\documentclass[aps,prd,reprint,superscriptaddress,longbibliography]{revtex4-2}

\usepackage[colorlinks=true,linkcolor=blue,citecolor=blue,urlcolor=blue]{hyperref}
\usepackage{microtype}
\usepackage{graphicx}
\usepackage{amsfonts,amsthm,amsmath,amssymb,bm}
\usepackage{booktabs}
\usepackage{placeins}
\usepackage{xcolor}
\usepackage{orcidlink}

\graphicspath{{figs/}}

\newcommand{\Tr}{\mathrm{Tr}}
\newcommand{\sgn}{\operatorname{sgn}}
\newcommand{\nb}{\bar n}
\newcommand{\hh}{\hat h}
\newcommand{\Jeff}{\mathcal{J}_{\rm eff}}
\newcommand{\Jpeff}{\mathcal{J}'_{\rm eff}}

\begin{document}

\title{\texorpdfstring{$U(1)$}{U(1)}-Symmetric Spin SYK Models: Ordering, Chaos, and Transport}

\author{Shiyong Guo${}^{\orcidlink{0009-0007-8775-8242}}$}
\affiliation{Department of Physics, Brandeis University, Waltham, Massachusetts, 02453}

\author{Enrico Rinaldi${}^{\orcidlink{0000-0003-4134-809X}}$}
\affiliation{Quantinuum, Partnership House, London, UK}

\author{Jad C.~Halimeh${}^{\orcidlink{0000-0002-0659-7990}}$}
\email{jad.halimeh@lmu.de}
\affiliation{Department of Physics and Arnold Sommerfeld Center for Theoretical Physics (ASC), Ludwig Maximilian University of Munich, 80333 Munich, Germany}
\affiliation{Max Planck Institute of Quantum Optics, 85748 Garching, Germany}
\affiliation{Munich Center for Quantum Science and Technology (MCQST), 80799 Munich, Germany}

\author{Brian Swingle${}^{\orcidlink{https://orcid.org/0000-0003-0334-3108}}$}
\email{bswingle@brandeis.edu}
\affiliation{Department of Physics, Brandeis University, Waltham, Massachusetts, 02453}

\begin{abstract}
The Sachdev--Ye--Kitaev (SYK) model provides an analytically controlled setting for strongly interacting quantum dynamics and holography, but its fermionic formulation leads to nonlocal encoding overheads on qubit-based quantum processors. We study a $U(1)$-symmetric spin analogue of the complex SYK model, constructed from random charge-conserving $q$-spin interactions that are native to a qubit Hilbert space. In the large-$N$, large-$q$ limit, the model reproduces the leading saddle-point structure of complex fermionic SYK under the identification of chemical potential with magnetic field and charge density with magnetization. We investigate the finite-$N$, finite-$q$ regime using exact diagonalization. A $U(1)$-invariant ordering susceptibility increases with system size for $q=2,4$ but decreases for $q=6$, while bulk-spectrum level statistics are consistent with the Gaussian orthogonal ensemble at half filling and the Gaussian unitary ensemble in the adjacent magnetization sector. We further construct coupled-cluster chains and obtain domain-wall dynamics consistent with diffusive magnetization transport in a small-cluster regime accessible to matrix-product-state simulations. These results establish analytical benchmarks and finite-size signatures of ordering, quantum chaos, and magnetization transport in $U(1)$-symmetric spin-SYK models, and they motivate quantum simulation of the low-temperature, large-system regime, which remains difficult for classical methods.
\end{abstract}

\maketitle

%%%%%%%%%%%%%%%%%%%%%%%%%%%%%%%%%%%%%%%%%%%%%%%%%%%%%%%%%%%%
\section{Introduction}
\label{sec:introduction}

The Sachdev--Ye--Kitaev (SYK) model provides a paradigmatic setting in which strongly interacting quantum dynamics, many-body quantum chaos, and aspects of holography can be studied analytically \cite{AddSachdevYe1993,AddMaldacenaStanford2016,Chowdhury2022Sep}. In its complex-fermion formulation, the model consists of $N$ fermionic degrees of freedom with all-to-all random $q$-body interactions. At large $N$, its correlation functions and thermodynamics are governed by self-consistent saddle-point equations, while at large $q$ these equations simplify further \cite{AddMaldacenaStanford2016,AddGuKitaevSachdev2020}. Closely related complex-SYK models display a conserved $U(1)$ charge, nontrivial thermoelectric transport, and thermodynamic structures with direct analogues in charged black-hole systems \cite{AddDavison2017,AddGuKitaevSachdev2020,Louw2023Feb,Louw2023Oct}. The Schwarzian mode, a low-energy collective mode associated with approximate time-reparametrization symmetry, and the maximal-chaos bound provide the standard bridge to nearly-AdS$_2$ quantum gravity \cite{AddMaldacenaShenkerStanford2016,AddMaldacenaStanfordYang2016,Sarosi:2018UG}.

This analytical control also makes SYK an attractive target for quantum simulation \cite{AddBabbush2019,AddAsaduzzaman2024,AddGranet2025,GarciaAlvarez2017,Luo2019NMR,Jafferis2022,Uhrich2023}. At the same time, its fermionic formulation creates a practical obstacle on qubit-based hardware: standard fermion-to-qubit encodings convert generic fermionic interaction terms into Pauli strings whose support can grow with system size, contributing substantially to circuit depth and gate cost \cite{AddBabbush2019,AddAsaduzzaman2024}. For complex SYK, each $q$-body term together with its conjugate maps under the Jordan--Wigner encoding to $2^q$ Pauli strings of weight up to $O(N)$, reduced to $O(q\log N)$ by the Bravyi--Kitaev encoding \cite{BravyiKitaev2002,Maskara2025}. A spin or hard-core-boson formulation avoids this particular nonlocal encoding, since each term of the spin model introduced below expands into the same number of Pauli strings, all of weight exactly $q$. The key question is whether the SYK-like physics of interest remains accessible in the spin model.

Here we study a $U(1)$-symmetric spin analogue of complex SYK. Spin and hard-core-boson relatives of SYK have a substantial history: hard-core bosonic all-to-all models, including the $q=4$ member of the family studied here, display glassy behavior, while more recent spin-SYK constructions exhibit strong quantum chaos and SYK-like thermodynamic behavior in appropriate regimes \cite{AddFuSachdev2016,AddHanadaSpin2024,Swingle2024Mar,Jia2024BosonicFlux,HanadaVanLeuven2026,BerkoozNarayanSimon2018,Watanabe2024}. In the model we consider, each cluster contains $N$ spin-$1/2$ degrees of freedom with random charge-conserving $q$-body interactions, and multiple clusters can be coupled into a one-dimensional chain. The conserved quantity is total magnetization, which plays the role of fermionic charge. This construction is useful for three complementary reasons. First, it admits analytically controlled limits connected to complex SYK. Second, at finite $N$ and $q$ it supports nontrivial many-body phenomena including spin ordering and quantum-chaotic spectra. Third, coupled clusters support conserved-charge transport and provide a setting in which entanglement growth can limit classical tensor-network methods. Related SYK-chain constructions display diffusive transport, spatial scrambling, and nontrivial entanglement dynamics \cite{AddGuQiStanford2017,AddGuLucasQi2017,AddDavison2017,SongJianBalents2017}.

The large-$N$ and large-$q$ limits serve as analytical benchmarks. At leading order in the large-$q$ expansion, the spin model inherits the saddle-point structure of complex SYK, with magnetic field playing the role of chemical potential and magnetization per spin corresponding to fermionic charge density measured relative to half filling.
The precise extent to which this correspondence survives at finite $N$ and finite $q$ is therefore an important question. In particular, glassiness can obstruct the self-averaging saddle-point description even when spectral diagnostics continue to indicate quantum chaos. The relation between SYK-like behavior and glassiness has been analyzed both in fermionic SYK and in hard-core-boson/spin variants \cite{AddFuSachdev2016,AddGurAri2018,Swingle2024Mar,Baldwin2020Aug,Anschuetz2025Nonglassy,AnousHaehl2021}, and the crossover from spin-liquid to spin-glass behavior has been studied in detail in random quantum Heisenberg magnets \cite{GeorgesParcolletSachdev2000,GeorgesParcolletSachdev2001,ChristosHaehlSachdev2022}.

We address this finite-size regime using exact diagonalization. We use a $U(1)$-invariant Edwards--Anderson (spin-glass) susceptibility that remains nonzero in finite systems even though $U(1)$ symmetry forces transverse one-point functions to vanish. Applied to the ground state, accessible system sizes show qualitatively different finite-size trends between $q=2,4$ and $q=6$. We separately analyze bulk-spectrum level statistics and the spectral form factor and find random-matrix behavior consistent with the Gaussian orthogonal ensemble at half filling and the Gaussian unitary ensemble in the adjacent magnetization sector. Adjacent-gap ratios and spectral form factors are standard finite-size probes of random-matrix universality in chaotic many-body systems and have been used extensively in dense and sparse SYK models \cite{AddAtas2013,AddCotler2017,AddGarciaGarcia2021,AddOrman2025}. Because the ordering diagnostic is evaluated in the ground state while the level statistics probe the bulk of the spectrum, these results should be viewed as complementary.

We then turn to coupled clusters. A local magnetization imbalance can spread from cluster to cluster while the total $U(1)$ charge remains conserved. Matrix-product-state simulations of a small-cluster regime show domain-wall broadening consistent with diffusive transport, in the spirit of transport studies in generalized and coupled SYK models \cite{AddGuQiStanford2017,AddDavison2017,Zanoci2022Jun}. We treat this transport calculation as a proof of principle that motivates further study of the low-temperature regime, where, by the correspondence with complex SYK, one expects Planckian dynamics: relaxation and scrambling times set by $\beta$, with the Lyapunov exponent approaching the chaos bound $2\pi T$ \cite{AddMaldacenaShenkerStanford2016,AddMaldacenaStanford2016,HartnollMackenzie2022,PatelSachdev2019,Bhattacharya2017}. Establishing this for the spin model requires an analysis of its own retarded ladder kernel, which we do not carry out here. We argue that quantum simulation is a promising route to probing this low-temperature regime.

The remainder of the paper is organized as follows. Section~\ref{sec:model} defines the spin-SYK model and its coupled-chain extension. Section~\ref{sec:limits} summarizes the controlled large-$N$ and large-$q$ limits. Section~\ref{sec:numerics} presents exact-diagonalization diagnostics of ordering and quantum chaos together with tensor-network results for magnetization transport. We conclude in Sec.~\ref{sec:outlook} with implications for future analytical, numerical, and quantum-simulation studies.

%%%%%%%%%%%%%%%%%%%%%%%%%%%%%%%%%%%%%%%%%%%%%%%%%%%%%%
\section{Model}
\label{sec:model}

We consider $L$ clusters, each containing $N$ spin-$1/2$ degrees of freedom. We use $\sigma^\pm = \frac{\sigma^x \pm i\sigma^y}{2}$, where $\sigma^{x,y,z}$ are Pauli matrices.

\subsection{Single-cluster Hamiltonian}

For even $q$, the Hamiltonian of cluster $i$ is
\begin{equation}
    H_i =
    J
    \sum_{\substack{\bm{\mu}\\ \bm{\nu}}}
    X^{(i)\bm{\mu}}_{\phantom{(i)}\bm{\nu}}\,
    \sigma^+_{\mu_1;i}\cdots \sigma^+_{\mu_{q/2};i}
    \sigma^-_{\nu_{q/2};i}\cdots \sigma^-_{\nu_1;i},
    \label{eq:single_cluster}
\end{equation}
where $X^{(i)\bm{\mu}}_{\phantom{(i)}\bm{\nu}}$ are random complex couplings, whose distribution and Hermiticity constraint are specified below, and $\bm{\mu}=(\mu_1,\ldots,\mu_{q/2})$ and
$\bm{\nu}=(\nu_1,\ldots,\nu_{q/2})$ are ordered tuples,
\begin{equation}
    1\le \mu_1<\cdots<\mu_{q/2}\le N,
    \qquad
    1\le \nu_1<\cdots<\nu_{q/2}\le N.
\end{equation}
We impose the Hermiticity condition
\begin{equation}
    X^{(i)\bm{\mu}}_{\phantom{(i)}\bm{\nu}}
    =
    \left(
    X^{(i)\bm{\nu}}_{\phantom{(i)}\bm{\mu}}
    \right)^* 
    \label{eq:hermiticity_X}
\end{equation}
and demand that $\bm{\mu} \cap \bm{\nu} = \emptyset$ so that every spin operator acts on a distinct spin. In the large $N$ limit, the terms with $\bm{\mu} \cap \bm{\nu} \neq \emptyset$ are a vanishing fraction of all the terms, but they significantly affect the finite $N$ physics if present. Forbidding them enhances the symmetry of the model (Sec.~\ref{sec:chaos}), ensures that every term in $H$ is exactly $q$-body, and removes terms analogous to chemical potential disorder that complicate transport (Sec.~\ref{sec:transport}). For $q=4$, Eq.~\eqref{eq:single_cluster} with this restriction is the hard-core-boson SYK model of Ref.~\cite{AddFuSachdev2016}, where pair hopping was likewise restricted to distinct sites in order to preserve particle--hole symmetry. 

The number of ordered tuple pairs is
\begin{equation}
    \binom{N}{q/2} \binom{N-q/2}{q/2}
    \sim
    \left[
    \frac{N^{q/2}}{(q/2)!}
    \right]^2
\end{equation}
for fixed $q$ and large $N$. The couplings $X^{(i)\bm{\mu}}_{\phantom{(i)}\bm{\nu}}$ for distinct unordered pairs $\{\bm{\mu},\bm{\nu}\}$ are independent proper complex Gaussians, with the partner entry fixed by Eq.~\eqref{eq:hermiticity_X}, zero mean, and variance
\begin{equation}
    \overline{|X|^2}
    =
    \frac{2^{q-1}N}{q^2}
    \left[
    \frac{N^{q/2}}{(q/2)!}
    \right]^{-2}.
    \label{eq:X_variance}
\end{equation}
These choices simplify the large-$N$ theory as explained in Section~\ref{sec:limits}.

The conserved cluster magnetization is
\begin{equation}
    M_i=\frac{1}{2}\sum_{\mu=1}^{N}\sigma^z_{\mu i},
    \label{eq:Mi}
\end{equation}
with
\begin{equation}
    [M_i,\sigma^\pm_{\mu i}]=\pm \sigma^\pm_{\mu i},
    \qquad
    [M_i,H_i]=0.
\end{equation}

\subsection{Coupled-cluster Hamiltonian}

We couple neighboring clusters by transferring $q'/2$ units of magnetization from cluster $i$ to cluster $i+1$,
\begin{equation}
\begin{aligned}
    H_{i\rightarrow i+1}
    &=J'\sum_{\substack{\bm{\mu}\\ \bm{\nu}}}
    Y^{(i\rightarrow i+1)\bm{\mu}}_{\phantom{(i\rightarrow i+1)}\bm{\nu}}
    \sigma^+_{\mu_1;i+1}\cdots \sigma^+_{\mu_{q'/2};i+1}
    \\
    &\qquad\times
    \sigma^-_{\nu_{q'/2};i}\cdots \sigma^-_{\nu_1;i}.
\end{aligned}
    \label{eq:intercluster}
\end{equation}
Note that no overlap condition is needed here since the spins in $\bm{\mu}$ and $\bm{\nu} $ reside in different clusters. The full Hamiltonian is
\begin{equation}
    H=\sum_i H_i+\sum_i
    \left(
        H_{i\rightarrow i+1}
        +H_{i\rightarrow i+1}^\dagger
    \right).
    \label{eq:fullH}
\end{equation}
No Hermiticity constraint is imposed on the directed-bond couplings $Y$: Hermiticity is ensured by explicitly adding $H_{i\to i+1}^{\dagger}$, whose coefficients are the complex conjugates of those in $H_{i\to i+1}$.
The intercluster couplings $Y^{(i\rightarrow i+1)\bm{\mu}}_{\phantom{(i\rightarrow i+1)}\bm{\nu}}$ are independent proper complex Gaussians, independent of each other across tuple pairs and bonds and of the $X^{(i)}$, with zero mean and variance
\begin{equation}
    \overline{|Y|^2}
    =
    \frac{2^{q'-1}N}{q'^2}
    \left[
    \frac{N^{q'/2}}{(q'/2)!}
    \right]^{-2}.
    \label{eq:Y_variance}
\end{equation}
For $J'=0$, each $M_i$ is separately conserved. At nonzero intercluster coupling only the total magnetization
\begin{equation}
    M_{\rm tot}=\sum_i M_i
\end{equation}
generates a conserved $U(1)$ symmetry. In general, the $U(1)^L$ symmetry of the decoupled set of $L$ clusters is broken to $[U(1)\times\mathbb{Z}_k^L]/\mathbb{Z}_k\simeq U(1)\times\mathbb{Z}_k^{L-1}$ with $k=q'/2$, reflecting the fact that the inter-cluster coupling only changes the local charge in units of $k$: each $e^{2\pi i(M_i+N/2)/k}$ commutes with $H$, and the product of all of them is a global rotation. For $q'>2$, states with definite local charge residues evolve within the corresponding sector; thermalization, if it occurs, is constrained by these conserved residues. For the $q'=2$ chains studied numerically below, the residues are trivial. Using
\begin{equation}
    [M_i,H_{i\rightarrow i+1}]
    =-\frac{q'}{2}H_{i\rightarrow i+1},
\end{equation}
one obtains a lattice continuity equation,
\begin{equation}
    \partial_t M_i
    =-\left(I_{i,i+1}-I_{i-1,i}\right),
    \label{eq:continuity}
\end{equation}
with bond-current operator
\begin{equation}
    I_{i,i+1}
    =
    -\,\frac{i q'}{2}
    \left(
        H_{i\rightarrow i+1}
        -H_{i\rightarrow i+1}^{\dagger}
    \right).
    \label{eq:current}
\end{equation}
This conserved current makes the coupled model a natural setting in which to study magnetization transport.

%%%%%%%%%%%%%%%%%%%%%%%%%%%%%%%%%%%%%%%%%%%%%%%%%%%%%%
\section{Controlled limits}
\label{sec:limits}

At fixed $L$ and $q$, the large-$N$ limit admits a disorder-averaged saddle-point formulation in terms of collective two-time correlation functions. In complex-fermion SYK this yields self-consistent Schwinger--Dyson equations \cite{Chowdhury2022Sep,AddGuKitaevSachdev2020}. Spin and hard-core-boson variants require a different local functional because operators on distinct sites commute \cite{AddFuSachdev2016,Swingle2024Mar}. Below we explain why this difference disappears at leading large $q$, state the resulting thermodynamic and transport benchmarks, and retain the limitations relevant to the finite-size numerics. Intermediate derivations are collected in App.~\ref{app:saddle-details}; the conductivity calculation is given in App.~\ref{app:conductivity}.

A further simplification occurs when the large-$q$ limit is taken after the large-$N$ limit. The motivation for the present construction is that, to leading order in this regime, the spin model reproduces the structure of the $U(1)$-symmetric fermionic SYK saddle under the following mapping. Here $\mu$ is the fermionic chemical potential, $h$ is the field conjugate to spin magnetization, $Q$ is the fermion-number operator of one cluster, and $M$ denotes the corresponding cluster magnetization $M_i$ with the cluster index suppressed:
\begin{equation}
    \mu \leftrightarrow h,
    \qquad
    Q/N-\tfrac12\leftrightarrow M/N. 
    \label{eq:mapping}
\end{equation}
The local occupation dictionary $n_\mu=c_\mu^\dagger c_\mu\leftrightarrow(1+\sigma_\mu^z)/2$ gives $Q=\sum_\mu n_\mu\leftrightarrow M+N/2$. Thus half filling corresponds to equal up- and down-spin populations and zero magnetization. Measuring charge relative to half filling also makes charge reversal transparent: $Q\to N-Q$ corresponds to $M\to-M$. This is an operator dictionary; it does not imply equivalence of the interacting spin and fermion dynamics at finite $q$.

This connects the spin model to the thermodynamic structures and transport phenomenology discussed for complex SYK and coupled complex-SYK models \cite{AddDavison2017,AddGuKitaevSachdev2020,AddGuQiStanford2017,Louw2023Feb,Louw2023Oct,Louw2024Sep,Zanoci2022Jun}.

The analysis below provides a solvable benchmark at large $N$ and large $q$, with $L$ fixed. It does not resolve the finite-$N$, finite-$q$ chain at large $L$. We use the annealed disorder average, averaging the partition function rather than its logarithm; possible glassy phases therefore require a separate analysis.

%%%%

\subsection{Conventions}
\label{subsec:limits-conv}

We set $\hbar=k_B=1$ and write $\beta=1/T$. We work in the grand-canonical ensemble $e^{-\beta(H-hM_{\rm tot})}$ with
$M_{\rm tot}=\sum_i M_i$ and $M_i=\tfrac12\sum_\mu\sigma^z_{\mu i}$.
For $q'>2$, the unrestricted grand-canonical trace sums over the conserved local charge-residue sectors described in Sec.~\ref{sec:model}; no projection onto a fixed residue sector is imposed here. Equivalence to a fixed-sector calculation is not assumed.
Normalized traces are written $\langle\!\langle\cdot\rangle\!\rangle=\Tr[\cdot]/2^{NL}$.
Time ordering $\mathcal{T}$ is plain chronological ordering (no sign),
appropriate to operators that commute at distinct sites. The basic bilocal field, a correlation function depending on two imaginary times, at cluster $i$ is
\begin{align}
    G_i(\tau_1,\tau_2)
   & =\frac1N\sum_{\mu=1}^N
    \big\langle\mathcal{T}\,\sigma^+_{\mu i}(\tau_1)\,\sigma^-_{\mu i}(\tau_2)\big\rangle, \nonumber \\
    G_i(\tau_1,\tau_2)&=G_i(\tau_1+\beta,\tau_2),
    \label{eq:Gdef}
\end{align}
which is periodic in each argument. Its two equal-time limits are the
filling and its complement,
\begin{equation}
    G_i(\tau^+,\tau)=\nb= \tfrac12+m, \,
    G_i(\tau^-,\tau)=1-\nb, \,
    m = \frac{\langle M_i\rangle}{N},
\end{equation}
so that $4G_i(\tau^\pm,\tau)G_i(\tau,\tau^\pm)=4\nb(1-\nb)=1-4m^2$.
We write $\int\!\!\int\equiv\int_0^\beta d\tau_1\!\int_0^\beta d\tau_2$.

On a single site the algebra $\{\sigma^+,\sigma^-\}=1$, $(\sigma^\pm)^2=0$
is that of one fermion mode, $\sigma^+\to c^\dagger$, $\sigma^-\to c$;
distinct sites commute for spins and anticommute for fermions. This difference is further discussed in Sec.~\ref{subsec:limits-diff}.

\subsection{\texorpdfstring{Annealed large-$N$ action for one cluster}{Annealed large-N action for one cluster}}
\label{subsec:limits-action}

Averaging the random couplings produces an interaction between collective two-time fields. Replacing the sums over distinct indices by powers of these fields is accurate at leading large $N$, with relative counting corrections of order $q^2/N$ (App.~\ref{app:averaging}).

Imposing the collective-field constraint with a bilocal Lagrange multiplier $\Sigma$, which acts as a self-consistent kernel nonlocal in imaginary time, gives the single-spin description. The $N$
spins decouple and one obtains the annealed (replica-diagonal) representation, which can differ from the quenched average in a glassy regime \cite{Baldwin2020Aug},
\begin{equation}
    \overline{Z}=\int\!\mathcal{D}G\,\mathcal{D}\Sigma\;e^{-N S[G,\Sigma]},
\end{equation}
\begin{align}
    -S[G,\Sigma]
    &=\ln Z_{\rm loc}[\Sigma]
    +\int\!\!\int\Sigma(\tau_2,\tau_1)\,G(\tau_1,\tau_2) \nonumber \\
    &+\frac{J^2}{4q^2}\int\!\!\int\big[4G(\tau_1,\tau_2)G(\tau_2,\tau_1)\big]^{q/2}
\label{eq:action}
\end{align}
\begin{align}
    Z_{\rm loc}[\Sigma]
    &=\Tr_{\,\text{1 spin}}\;\mathcal{T}\exp\Big[
    \frac h2\int_0^\beta\! d\tau\,\sigma^z(\tau)
    \nonumber\\[-2pt]
    &\qquad-\int\!\!\int\Sigma(\tau_2,\tau_1)\,
    \sigma^+(\tau_1)\sigma^-(\tau_2)\Big].
    \label{eq:Zloc}
\end{align}
$Z_{\rm loc}$ is the partition function of a single spin-$1/2$ in a field $h$, with a self-interaction nonlocal in imaginary time. It plays the role of the fermionic determinant $\det(\partial_\tau+\mu-\Sigma)$ in complex SYK, but is not a Gaussian functional of the local fields. Evaluating this single-spin functional is the main obstacle to solving the saddle at finite $q$.

\paragraph{Saddle-point equations.}
Varying Eq.~\eqref{eq:action} (both orderings $(\tau_1,\tau_2)$ and
$(\tau_2,\tau_1)$ of the double integral contribute to $\delta/\delta G(\tau_1,\tau_2)$):
\begin{align}
    G(\tau_1,\tau_2)&=\big\langle\mathcal{T}\sigma^+(\tau_1)\sigma^-(\tau_2)\big\rangle_{\Sigma},
    \label{eq:EOM1}\\
    \Sigma(\tau_1,\tau_2)&=-\frac{J^2}{q}\,
    \big[4G(\tau_1,\tau_2)G(\tau_2,\tau_1)\big]^{q/2-1}\,G(\tau_1,\tau_2),
    \label{eq:EOM2}
\end{align}
where $\langle\cdot\rangle_\Sigma$ is the expectation in the single-spin
problem \eqref{eq:Zloc}. Equation~\eqref{eq:EOM2} is the melonic
self-energy; Eq.~\eqref{eq:EOM1} replaces the fermionic Dyson equation.

\subsection{Dictionary to complex SYK}
\label{subsec:limits-dict}

Let $G_f(\tau_1,\tau_2)\equiv\langle\mathcal{T}_F\,c^\dagger(\tau_1)c(\tau_2)\rangle$
with fermionic time ordering ($G_f$ is the transpose of the usual
$\mathcal{G}(\tau_1,\tau_2)=-\langle\mathcal{T}_Fc(\tau_1)c^\dagger(\tau_2)\rangle$,
i.e.\ $\mathcal{G}(\tau)=G_f(-\tau)$). On a single site the operator strings
are identical and only the ordering sign differs, so for the same local operator problem
(the same $\Sigma$ in Eq.~\eqref{eq:Zloc}) the two orderings are related by
\begin{equation}
\begin{aligned}
    G_f(\tau_1,\tau_2)
    &=\sgn(\tau_1-\tau_2)\,G(\tau_1,\tau_2),
    \\
    G_f(\tau_1,\tau_2)G_f(\tau_2,\tau_1)
    &=-\,G(\tau_1,\tau_2)G(\tau_2,\tau_1).
\end{aligned}
    \label{eq:dict}
\end{equation}
The interaction term of Eq.~\eqref{eq:action} therefore has the same form as the
complex-SYK interaction term, whose vertex is
$(-1)^{q/2}[\,G_f(\tau_1,\tau_2)G_f(\tau_2,\tau_1)]^{q/2}
=[G(\tau_1,\tau_2)G(\tau_2,\tau_1)]^{q/2}$ up to the same normalization. Thermodynamically,
$M=Q-N/2$ gives
\begin{equation}
\begin{aligned}
    e^{-\beta(H-hM)}&=e^{-\beta hN/2}\,e^{-\beta(H-hQ)},
    \\
    \mu&=h,\qquad m=\frac{\langle Q\rangle}{N}-\frac12\equiv\mathcal{Q},
\end{aligned}
    \label{eq:mapping-fixed}
\end{equation}
so the charge that maps onto $M/N$ is the charge density \emph{measured from
half filling}. Equation~\eqref{eq:dict} compares ordering conventions for the
same single-site operator problem; it does not assert that the interacting
spin and fermion correlators coincide at finite $q$, which they do not
(Sec.~\ref{subsec:limits-diff}).

\subsection{Where spins and fermions differ: the local functional}
\label{subsec:limits-diff}

To explain the difference with fermionic SYK, begin by expanding $\ln Z_{\rm loc}[\Sigma]$ in powers of $\Sigma$. The $n$-th order term
involves the $2n$-point function
$\langle\mathcal{T}\sigma^+(\tau_1)\sigma^-(\tau_2)\cdots\sigma^+(\tau_{2n-1})\sigma^-(\tau_{2n})\rangle_h$
of a free spin. For the fermion the analogous object is Wick-factorizable; for
the spin it is not, but on a single site the two differ only by the sign of the
permutation that chronologically orders the string.

Now, through second order in $\Sigma$, the spin and fermion local functionals
coincide under the dressing
$\Sigma_f(\tau_1,\tau_2)=\sgn(\tau_2-\tau_1)\,\Sigma(\tau_1,\tau_2)$.
Equivalently, the linear response of Eq.~\eqref{eq:EOM1} is the fermionic
Dyson equation:
\begin{equation}
\begin{aligned}
    \delta G_f(\tau_1,\tau_2)
    &=\int\!\!\int G^0_f(\tau_1,\tau_4)\,\Sigma_f(\tau_4,\tau_3)
    \\
    &\qquad\times G^0_f(\tau_3,\tau_2)+O(\Sigma^2).
\end{aligned}
    \label{eq:linDyson}
\end{equation}
Here $G_f^0$ denotes the free single-site fermionic propagator used in the expansion about $\Sigma=0$.

The equality through quadratic order follows from the four-point functions of a single free spin. At cubic order, six-point functions contain chronological-ordering signs that cannot be absorbed into the same pairwise dressing. Explicit orderings establishing both statements are given in App.~\ref{app:local-signs}.

Hence, the replica-diagonal large-$N$ saddles of the spin model and of complex
SYK share the same interaction term and the same quadratic local kernel, and
differ starting at the cubic ($6$-point) local vertex. This statement is the precise content of ``the spin model
reproduces the saddle of complex SYK'', but these results do not guarantee any equivalence at finite $q$ and do not address any possible glassy structure.

\subsection{\texorpdfstring{Large $q$}{Large q}}
\label{subsec:limits-largeq}

From Eq.~\eqref{eq:EOM2}, $\Sigma=O(J^2/q)$ at fixed $J$ and fixed $\beta$.
Only the quadratic part of $\ln Z_{\rm loc}$ contributes to the saddle at
leading order in $1/q$, so the spin problem reduces to the
fermionic one; the cubic vertex enters at relative order $1/q$. The large-$q$ expansion is taken at fixed dimensionless inverse temperature $\beta\Jeff$ and is not uniform as $T/\Jeff\to0$.

\paragraph{Effective field.}
Let $\hh$ be the field for which a \emph{free} spin has filling $\nb$,
\begin{equation}
    \nb=\frac{1}{1+e^{-\beta\hh}}
    \quad\Longleftrightarrow\quad
    \hh=\frac1\beta\ln\frac{1+2m}{1-2m},
\end{equation}
and let $G^0$ be its propagator,
$G^0(\tau)=e^{-\hh\tau}\big[\nb\,\theta(\tau)+(1-\nb)\,\theta(-\tau)\big]$
(periodic), so that $G^0_f=\sgn\cdot G^0$ is antiperiodic and satisfies
$(\partial_\tau+\hh)G^0_f=\delta(\tau)$. The physical $h$ differs from $\hh$
by an interaction correction of order $1/q$; it drops out of the equations
below away from coincident times and is fixed by the equation of state
(Sec.~\ref{subsec:limits-thermo}).

\paragraph{Ansatz and Liouville equation.}
Write
\begin{equation}
\begin{aligned}
    G(\tau_1,\tau_2)
    &=G^0(\tau_1,\tau_2)
    \Big[1+\frac{g(\tau_1,\tau_2)}{q}+O(q^{-2})\Big],
    \\
    g(\tau,\tau)&=0 .
\end{aligned}
\end{equation}
The symmetric correction $g_s(\tau_1,\tau_2)=[g(\tau_1,\tau_2)+g(\tau_2,\tau_1)]/2$ obeys a closed differential equation. The free-spin product $G^0(\tau)G^0(-\tau)=\nb(1-\nb)$ fixes its filling dependence, while periodicity removes the antisymmetric zero mode (App.~\ref{app:largeq-steps}). In equilibrium, $g_s$ depends only on the time difference, and the result is

\begin{equation}
\;
    \partial_\tau^2\,g_s(\tau)=2\,\Jeff^2\,e^{\,g_s(\tau)},
    \qquad
    \Jeff^2=\frac{J^2}{2}\,\big[1-4m^2\big]^{\,q/2-1}
\;
\label{eq:liouville}
\end{equation}
with the standard solution
\begin{equation}
\begin{aligned}
    e^{g_s(\tau)}&=\left[\frac{\cos(\pi v/2)}
    {\cos\!\big(\pi v(\tfrac12-\tfrac{\tau}{\beta})\big)}\right]^{2},
    \\
    \beta\Jeff&=\frac{\pi v}{\cos(\pi v/2)},\qquad 0\le v<1 .
\end{aligned}
    \label{eq:thermal-solution}
\end{equation}
This is the large-$q$ complex-SYK Liouville equation under
Eq.~\eqref{eq:mapping-fixed}: the filling enters only through the
$U(1)$-invariant $1-4m^2=1-4\mathcal{Q}^2$, with the same exponent $q/2-1$,
because that exponent is produced entirely by the equal-time single-site
factors, where the algebras agree. As in complex SYK, keeping $\Jeff=O(1)$ as
$q\to\infty$ at $m\neq0$ requires either scaling $m=O(q^{-1/2})$ or treating
$\Jeff$ rather than $J$ as the fixed parameter.

An independent short-time commutator check verifies the factor $1/2$ in $\Jeff^2$ (App.~\ref{app:single-shorttime}).

\subsection{Thermodynamics and equation of state}
\label{subsec:limits-thermo}

It is cleanest to work first at fixed magnetization.  Define
$A(m)=1-4m^2$ and $\nb=1/2+m$.  In terms of the parameter $v$ in
Eq.~\eqref{eq:thermal-solution}, the universal one-complex-mode function is
\begin{equation}
\begin{aligned}
    \Phi_{\rm c}(x)
    &=2\pi v\tan\frac{\pi v}{2}-\frac{\pi^2v^2}{2},
    \\
    x&=\beta\Jeff=\pi v\sec\frac{\pi v}{2}.
\end{aligned}
    \label{eq:Phi-complex}
\end{equation}
It is twice the Majorana function in the same normalization, as required by
the two Euclidean channels of one complex mode.  The Helmholtz free energy per
spin is therefore
\begin{equation}
\begin{aligned}
    \beta f(m,T)
    ={}&\nb\ln\nb+(1-\nb)\ln(1-\nb)
    \\
    &-\frac{A(m)}{q^2}\,
    \Phi_{\rm c}\!\big(\beta\Jeff(m)\big)+O(q^{-3}).
\end{aligned}
    \label{eq:canonical-free-energy}
\end{equation}
The grand potential per spin is obtained only after the Legendre transform,
\begin{equation}
    \frac{\Omega(h,T)}{N}
    =\min_{-1/2\leq m\leq1/2}\big\{f(m,T)-hm\big\}.
    \label{eq:grand-potential}
\end{equation}
Equation~\eqref{eq:canonical-free-energy} follows by integrating the saddle-point energy from $\beta=0$ (App.~\ref{app:thermo-integration}). It agrees with the complex-SYK result under Eq.~\eqref{eq:mapping-fixed} \cite{Louw2023Feb}.

Differentiating Eq.~\eqref{eq:canonical-free-energy} gives the physical
equation of state
\begin{equation}
    h=2T\tanh^{-1}(2m)
    +\frac{4m}{q}\Jeff(m)\sin\frac{\pi v}{2}+O(q^{-2}).
    \label{eq:equation-of-state}
\end{equation}
Thus $\hh=2T\tanh^{-1}(2m)$ and $h-\hh=O(q^{-1})$.  The factor in the
correction follows from
$\partial_m\ln\Jeff=-2(q-2)m/A(m)$.  Defining the susceptibility per spin by
$\chi_m=(\partial m/\partial h)_T$, one obtains
\begin{equation}
\begin{aligned}
    \chi_m^{-1}
    &=\frac{4T}{A(m)}
    +\frac{4}{q}\frac{\partial}{\partial m}
    \left[m\Jeff(m)\sin\frac{\pi v}{2}\right]+O(q^{-2}),
    \\
    \chi_m^{-1}\big|_{m=0}
    &=4T+\frac{4\Jeff(0)}{q}\sin\frac{\pi v}{2}+O(q^{-2}),
\end{aligned}
    \label{eq:susceptibility}
\end{equation}
where the $m$ dependence of $v$ is fixed implicitly by
Eq.~\eqref{eq:thermal-solution}.

\subsection{Coupled chain}
\label{subsec:limits-chain}

Averaging the directed-bond couplings adds a nearest-neighbor interaction between the collective fields $G_i$. The action and self-consistency equations are given in App.~\ref{app:chain-action}. The two hopping directions fix the relative normalization of the bond and intracluster terms. Distinct indices within each bond tuple still give relative $O(q'^2/N)$ counting corrections; these vanish for $q'=2$.

The bilocals $G_i$ are neutral, so constant $U(1)$ rotations act on them
trivially; the hydrodynamics is instead controlled by time-dependent rotations
$G_i(\tau_1,\tau_2)\to e^{i[\phi_i(\tau_1)-\phi_i(\tau_2)]}G_i(\tau_1,\tau_2)$.
The intracluster interaction term of Eq.~\eqref{eq:chain-action} is invariant
for arbitrary $\phi_i(\tau)$, while the bond term is invariant only when
$\phi_{i+1}(\tau)-\phi_i(\tau)$ is independent of $\tau$. In
$\ln Z_{\rm loc}[\Sigma_i]$, $\partial_\tau\phi_i$ enters as an
imaginary-time-dependent shift of $h$, which sets the compressibility $\chi$;
the bond term penalizes relative phase twists between neighboring clusters and
sets the conductivity $\sigma$. A uniform rotation $\phi_i(\tau)=\phi(\tau)$
leaves both interaction terms invariant, reflecting conservation of
$M_{\rm tot}$. These phase fluctuations describe the conserved-magnetization hydrodynamics. The scalar relation $D=\sigma/\chi$ applies when the magnetization mode decouples from energy, as specified below.

\paragraph{Homogeneous saddle and its large-$q$ limit.}
For $G_i=G$ the bond term acts like an additional single-cluster interaction.
With $q'=q$ and the ansatz of Sec.~\ref{subsec:limits-largeq},
\begin{equation}
\begin{aligned}
    \partial_\tau^2 g_s&=2\big(\Jeff^2+(\Jpeff)^2\big)e^{g_s},
    \\
    (\Jpeff)^2&=J'^2\big[1-4m^2\big]^{q/2-1},
\end{aligned}
\end{equation}
so the chain's equilibrium saddle is the single-cluster one with
$\Jeff^2\to\mathcal{J}_{\rm ch}^2\equiv\Jeff^2+(\Jpeff)^2$ and
$\beta\mathcal{J}_{\rm ch}=\pi v\sec(\pi v/2)$.  The equation of state and
$\chi_m$ follow from Eqs.~\eqref{eq:equation-of-state} and
\eqref{eq:susceptibility} with $\Jeff\to\mathcal{J}_{\rm ch}$.  For
$q'\neq q$ the bond contributes $e^{(q'/q)g_s}$ instead of $e^{g_s}$.  The numerics of
Sec.~\ref{sec:transport} use $q'=q=2$; they share the same Hamiltonian convention but do not
lie in the asymptotic large-$q$ regime.

\paragraph{Transport.}
Magnetization diffusion follows from the quadratic fluctuations of the phase
modes $\phi_i$ around the homogeneous saddle, or equivalently from the Kubo
formula with the current of Eq.~\eqref{eq:current}. Energy is also conserved,
so in general magnetization and energy diffuse as a coupled pair
\cite{AddDavison2017}. At half filling the antiunitary symmetry $PK$ of
Sec.~\ref{sec:chaos}, which extends to the chain because it maps each
$H_{i\to i+1}$ to $H_{i\to i+1}^\dagger$, sends $M_i\to-M_i$ and $H\to H$, so the static and Onsager
cross terms vanish and the two modes decouple at any temperature; there
$D=\sigma/\chi$ below is the magnetization diffusion constant. Away from half
filling the modes generally mix and the diffusion matrix is required
\cite{AddDavison2017}. At $\beta=0$ the static cross-susceptibility vanishes,
because $H$ has no diagonal elements in the $\sigma^z$ basis, so $\sigma/\chi$
equals the diagonal element $D_{MM}$; the Onsager cross term need not vanish, and
we do not compute it. With this qualification,
\begin{equation}
\begin{aligned}
    D&=\frac{\sigma}{\chi},\qquad
    \chi=\frac{\partial\langle M_i\rangle}{\partial h},
    \\
    \sigma&=\lim_{\omega\to0}\frac{1}{\omega}\,\mathrm{Im}\,
    \frac{1}{L}\big\langle I_{\rm tot};I_{\rm tot}\big\rangle^{R}(\omega),
\end{aligned}
\end{equation}
with lattice spacing set to one and
$\langle A;B\rangle^{R}(t)=i\theta(t)\langle[A(t),B]\rangle$, where $I_{\rm tot}=\sum_i I_{i,i+1}$. At leading order in $1/N$ the
total-current correlator per bond reduces to a single bond bubble built from $G$
(App.~\ref{app:conductivity}), although single-bond and cross-bond correlators
separately do not.  For the controlled $q'=q$ chain, the resulting dc
conductivity is (App.~\ref{app:conductivity}; for related calculations in fermionic SYK chains see Refs.~\cite{Jha2024Jul,Louw2024Sep})
\begin{equation}
    \frac{\sigma_{\rm dc}}{N}
    =\frac{\pi v}{4}A(m)\,
    \frac{(\Jpeff)^2}{\mathcal{J}_{\rm ch}^2}.
    \label{eq:largeq-conductivity}
\end{equation}
This differs from the large-$q$ complex-SYK lattices of Ref.~\cite{JhaLouw2023}, in which intersite hopping moves $r=O(q^0)$ fermions with a variance carrying an additional factor $1/q$, so that local charge densities remain constant at leading order in $1/q$. Here $q'=q$: each bond term transfers $q/2$ units of magnetization and enters the saddle at the same order as the intracluster interaction, $\mathcal{J}_{\rm ch}^2=\Jeff^2+(\Jpeff)^2$, so magnetization is exchanged between neighboring clusters on times of order $1/\mathcal{J}_{\rm ch}$.
Since $\chi=N\chi_m$, the Einstein relation gives
\begin{equation}
\begin{aligned}
    D
    &=\frac{\pi v}{4}A(m)\,
    \frac{(\Jpeff)^2}{\mathcal{J}_{\rm ch}^2}\,\chi_m^{-1}
    \\
    &=\frac{J'^2}{\mathcal{J}_{\rm ch}}\,
    \mathcal{F}_{q'=q}(\beta\mathcal{J}_{\rm ch},m),
\end{aligned}
    \label{eq:largeq-diffusion}
\end{equation}
where
\begin{equation}
    \mathcal{F}_{q'=q}
    =\frac{\pi v}{4}A(m)^{q/2}
    \frac{\chi_m^{-1}}{\mathcal{J}_{\rm ch}}.
    \label{eq:transport-function}
\end{equation}
At half filling this reduces to the explicit form
\begin{equation}
    \mathcal{F}_{q'=q}(m=0)
    \simeq\pi v\left[\frac{T}{\mathcal{J}_{\rm ch}}
    +\frac{1}{q}\sin\frac{\pi v}{2}\right].
    \label{eq:transport-function-half}
\end{equation}
This combines the leading-order conductivity with the $1/q$ correction to
$\chi_m$. The conductivity has its own $O(1/q)$ correction, which we have not
computed, so Eq.~\eqref{eq:transport-function-half} is a leading result with a
partial subleading correction. The expansion is controlled at fixed
$\beta\mathcal{J}_{\rm ch}$ as $q\to\infty$, not uniformly as $T\to0$: for
$T\lesssim\mathcal{J}_{\rm ch}/q$ the $1/q$ term in $\chi_m^{-1}$ is comparable to
the leading one.
Equations~\eqref{eq:largeq-conductivity}--\eqref{eq:transport-function-half}
provide the analytic benchmark for Sec.~\ref{sec:transport}. For $\beta\to0$ at fixed $m$,
$\pi v\simeq\beta\mathcal{J}_{\rm ch}$ and $\chi_m^{-1}\simeq4T/A(m)$, so
$\mathcal{F}_{q'=q}\to A(m)^{q/2-1}$; at $q=q'=2$ this gives
$D=J'^2/\mathcal{J}_{\rm ch}$ with $\mathcal{J}_{\rm ch}^2=J^2/2+J'^2$ (for $N\to\infty$; see Sec.~\ref{sec:transport} for finite $N$), independent of $m$ (for $m\neq0$ as the diagonal element $D_{MM}$).
The numerics of Sec.~\ref{sec:transport} have $q=2$, outside the controlled large-$q$
regime, so the comparison there is a consistency check rather than a controlled test.

%%%%%%%%%%%%%%%%%%%%%%%%%%%%%%%%%%%%%%%%%%%%%%%%%%%%%%%%%%%
\section{Numerical study}
\label{sec:numerics}

We now investigate the finite-$N$, finite-$q$ regime using exact diagonalization for a single cluster and matrix-product-state time evolution for coupled clusters.

For $L=1$, we work in the half-filling magnetization sector, whose Hilbert-space dimension is
\begin{equation}
    \mathcal{D}=\binom{N}{N/2}.
\end{equation}
The Hamiltonian is represented sparsely in the fixed-magnetization basis. We use Lanczos methods for the ground state, with system sizes up to $N=20$, and full diagonalization at $N=12$ for level statistics and the spectral form factor. Table~\ref{tab:numerics} lists the sizes, disorder realizations, and methods used for each result.

\begin{table}[t]
    \centering
    \caption{Numerical calculations. $R$ is the number of disorder realizations; for the chains, $L$ is the number of clusters and $\chi_{\rm MPS}$ the MPS bond dimension.}
    \label{tab:numerics}
    \begin{tabular}{@{}lllll@{}}
        \toprule
        Result & Sizes & $\chi_{\rm MPS}$ & $R$ & Method \\
        \midrule
        Ordering, Fig.~\ref{fig:chi_xy} & $q=2,4$: $N=6$--$20$ & -- & $30$ & Lanczos \\
         & $q=6$: $N=8$--$16$ & -- & $30$ & \\
        Chaos, Fig.~\ref{fig:chaos}, & $N=12$, two sectors & -- & $100$ & full diag. \\
        \quad Tab.~\ref{tab:gapratio} & & & & \\
        Transport, Fig.~\ref{fig:transport} & $N=2$, $L=48$ & $128$ & $100$ & TEBD \\
         & $N=3$, $L=48$ & $128$ & $48$ & \\
         & $N=4$, $L=32$ & $512$ & $8$ & \\
        Weak walls & $N=3$, $L=24$ & $\leq512$ & $48$ & TEBD \\
        Free fermions & $N=2$--$4$, as above & -- & $200$ & exact \\
        \bottomrule
    \end{tabular}
\end{table}

For $L>1$, each cluster is treated as a single MPS site with local dimension $d=2^N$. Real-time evolution is performed with second-order TEBD as implemented in TeNPy \cite{Hauschild2018Oct}; we use standard MPS diagnostics and convergence checks with respect to time step and bond dimension \cite{Schollwoeck2011Jan}.

\subsection{Finite-size ordering diagnostic}
\label{sec:glass}

Diagnostics of ordering that spontaneously breaks a symmetry in SYK-like models require care because symmetric finite-size eigenstates need not exhibit nonzero one-point order parameters \cite{AddFuSachdev2016,AddGurAri2018,Swingle2024Mar}. For a single cluster at half filling (dropping the cluster index), the $U(1)$ symmetry forces $\langle\sigma^\pm_\mu\rangle=0$ in every eigenstate of definite magnetization, and the antiunitary symmetry $PK$ of the half-filled sector (Sec.~\ref{sec:chaos}), which maps $\sigma^z_\mu\to-\sigma^z_\mu$, forces $\langle\sigma^z_\mu\rangle=0$ in every nondegenerate eigenstate. Both the transverse and the longitudinal Edwards--Anderson one-point constructions therefore vanish identically at finite $N$. We instead use the $U(1)$-invariant ground-state correlations
\begin{equation}
    C_{\mu\nu}^{xy}
    =
    \langle\psi_0|\sigma_\mu^+\sigma_\nu^-|\psi_0\rangle,
    \qquad
    \chi_{\mathrm{EA}}^{xy,\mathrm{off}}
    =
    \frac{1}{N}
    \sum_{\mu\neq\nu}
    |C_{\mu\nu}^{xy}|^2,
    \label{eq:chi_sg}
\end{equation}
where $|\psi_0\rangle$ is the ground state in the half-filled sector.

Every term of $H$ moves spins between up and down, so frozen order in this model plausibly takes the form of frozen transverse moments $b_\mu=\langle\sigma^-_\mu\rangle = \frac{1}{2} \langle (\sigma_x - i \sigma_y)\rangle$ in a symmetry-broken pure state; for $q=2$ the corresponding classical energy $\sum_{\mu\nu}X_{\mu\nu}\,b_\mu^*b_\nu$ is that of an XY spin glass with random complex couplings. Restoring the symmetry by projecting onto fixed magnetization gives $C^{xy}_{\mu\nu}\simeq b_\mu^*b_\nu$, and hence
\begin{equation}
    \frac{\chi_{\mathrm{EA}}^{xy,\mathrm{off}}}{N}
    \simeq\frac{1}{N^2}\sum_{\mu\neq\nu}|b_\mu|^2|b_\nu|^2,
    \label{eq:chi_bound}
\end{equation}
which for classical moments frozen in the XY plane, $|b_\mu|=1/2$, equals $(1-1/N)/16\to1/16$. We use $1/16$ as a classical reference value; it is not a strict bound on finite-$N$ quantum states. In a paramagnet, by contrast, the correlation between two given spins is built from coupling chains connecting them. For $q=2$ a single coupling suffices, $C^{xy}_{\mu\nu}=O(N^{-1/2})$, and $\chi_{\mathrm{EA}}^{xy,\mathrm{off}}$ stays $O(1)$. For $q\geq4$ the disjointness condition forbids a single term from connecting $\mu$ to $\nu$; the shortest chains involve two couplings, with $\sim N^{q-1}$ such chains of random phase, each of size $\overline{|X|^2}\sim N^{1-q}$, giving $C^{xy}_{\mu\nu}=O(N^{-(q-1)/2})$ and $\chi_{\mathrm{EA}}^{xy,\mathrm{off}}\sim N^{2-q}$. This is the expectation from large-$N$ counting in a paramagnetic phase rather than a derived zero-temperature result. The same counting organizes single-instance expansions of SYK observables in powers of $N^{-1/2}$~\cite{Swingle2026Sep}. In either case $\chi_{\mathrm{EA}}^{xy,\mathrm{off}}/N\to0$. Because the model is all-to-all, the relevant distinction is this scaling with $N$ rather than spatial decay. With random, frustrated couplings, a nonzero limit of $\chi_{\mathrm{EA}}^{xy,\mathrm{off}}/N$ signals Edwards--Anderson order. Whether the frozen state also breaks replica symmetry cannot be decided from the ground state alone.

\begin{figure*}[t]
    \centering
    \includegraphics[width=0.8\textwidth]{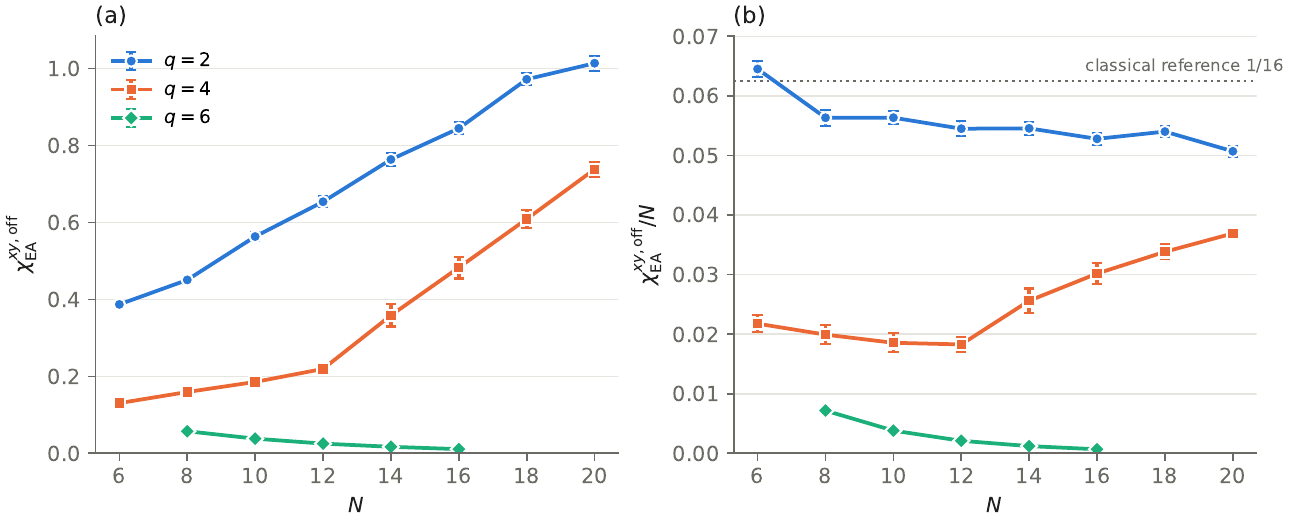}
    \caption{
    (a) $U(1)$-invariant susceptibility
    $\chi_{\mathrm{EA}}^{xy,\mathrm{off}}$ and (b) $\chi_{\mathrm{EA}}^{xy,\mathrm{off}}/N$ versus $N$ for
    $q=2,4,6$ in the half-filled ground state, averaged over $R=30$ disorder realizations
    (error bars: $\pm1$ standard error). The dotted line in (b) is the classical reference value $1/16$ of
    Eq.~\eqref{eq:chi_bound}.
    For $q=2$, $\chi_{\mathrm{EA}}^{xy,\mathrm{off}}$ grows linearly in $N$, with
    $\chi_{\mathrm{EA}}^{xy,\mathrm{off}}/N\approx0.05$; for $q=4$ it grows
    faster than linearly beyond $N\approx12$; for $q=6$ it decreases over the
    accessible sizes, all of which have $N<q^2$.
    }
    \label{fig:chi_xy}
\end{figure*}

Figure~\ref{fig:chi_xy} shows the results. For $q=2$, $\chi_{\mathrm{EA}}^{xy,\mathrm{off}}/N$ decreases only slowly, from $0.056$ at $N=8$ to $0.051\pm0.001$ at $N=20$; extrapolating linearly in $1/N$ or $1/\sqrt{N}$ gives $0.049\pm0.001$ or $0.043\pm0.003$, that is, $70$--$80\%$ of the classical reference value. These data are strong evidence for transverse order in the ground state. For $q=4$, $\chi_{\mathrm{EA}}^{xy,\mathrm{off}}/N\approx0.02$ and slowly decreasing up to $N=12$, then rises to $0.037\pm0.001$ at $N=20$. Even for $N\leq12$, $\chi_{\mathrm{EA}}^{xy,\mathrm{off}}$ itself grows, roughly as $N^{0.75}$, whereas the paramagnetic counting estimate predicts $N^{-2}$, so the $q=4$ ground state is strongly correlated before the upturn. The upturn is not driven by a few realizations: at $N=20$ the median of $\chi_{\mathrm{EA}}^{xy,\mathrm{off}}/N$ is $0.038$, and every realization exceeds the $N=12$ mean of $0.018$. The increments shrink with $N$, but the data have not saturated, so the large-$N$ value is not determined. These trends are consistent with the spin-glass order inferred for the same $q=4$ model in Ref.~\cite{AddFuSachdev2016} from the growth with $N$ of a low-frequency peak in the boson Green's function, a dynamical Edwards--Anderson diagnostic complementary to the static one used here. For $q=6$, $\chi_{\mathrm{EA}}^{xy,\mathrm{off}}/N$ decreases by a factor of about $0.55$ per $\Delta N=2$ over $8\leq N\leq16$, and no frozen order is visible at these sizes. The local exponent $d\ln\chi_{\mathrm{EA}}^{xy,\mathrm{off}}/d\ln N$ steepens from $-1.8$ to $-3.3$ over this range, qualitatively compatible with the heuristic counting estimate. These sizes do not establish convergence to an asymptotic $N^{-4}$ law. All accessible $q=6$ sizes have $N<q^2$, where the $O(q^2/N)$ corrections of Sec.~\ref{subsec:limits-action} are not small; this is also the scale at which the density of states of random $p$-local spin Hamiltonians crosses over from Gaussian to semicircular \cite{ErdosSchroder2014}. Moreover, the $q=4$ data show that finite-size trends can reverse within the accessible range. A similar $p$-dependence appears in exact diagonalization of random all-to-all $p$-spin models without $U(1)$ symmetry, where an Edwards--Anderson order parameter grows with $N$ for $p\leq5$ but decreases at $p=6$, with the crossover controlled by $N/p^2$ \cite{WadashimaMotome2025}. Therefore, while the $q=6$ data are consistent with a paramagnetic state, they leave room for changes in behavior at larger $N$. It is known that the quenched and annealed free energies must eventually differ in any bosonic spin model with fixed $q$, even in a paramagnetic phase~\cite{Baldwin2020Aug}. However, this does not mean the model is glassy, and there is evidence from other spin models that sufficiently large but finite $q$ can give a non-glassy state~\cite{Swingle2024Mar}. Charge-conserving qubit models built from Fock-space fluxes, which are not $p$-local, avoid low-temperature ordering altogether \cite{Jia2024BosonicFlux}. 

Distinguishing a replica-symmetric frozen state from a spin glass with many competing states would require overlaps between low-lying states or finite-temperature data, and resolving the large-$N$ behavior for $q\geq4$ requires larger $N$.

\subsection{Quantum chaos diagnostics}
\label{sec:chaos}

We next analyze spectral statistics at $N=12$, in the half-filled sector and, for comparison, in the sector with one additional up spin, averaging over $R=100$ disorder realizations.

\begin{figure*}[!t]
    \centering
    \includegraphics[width=\linewidth]{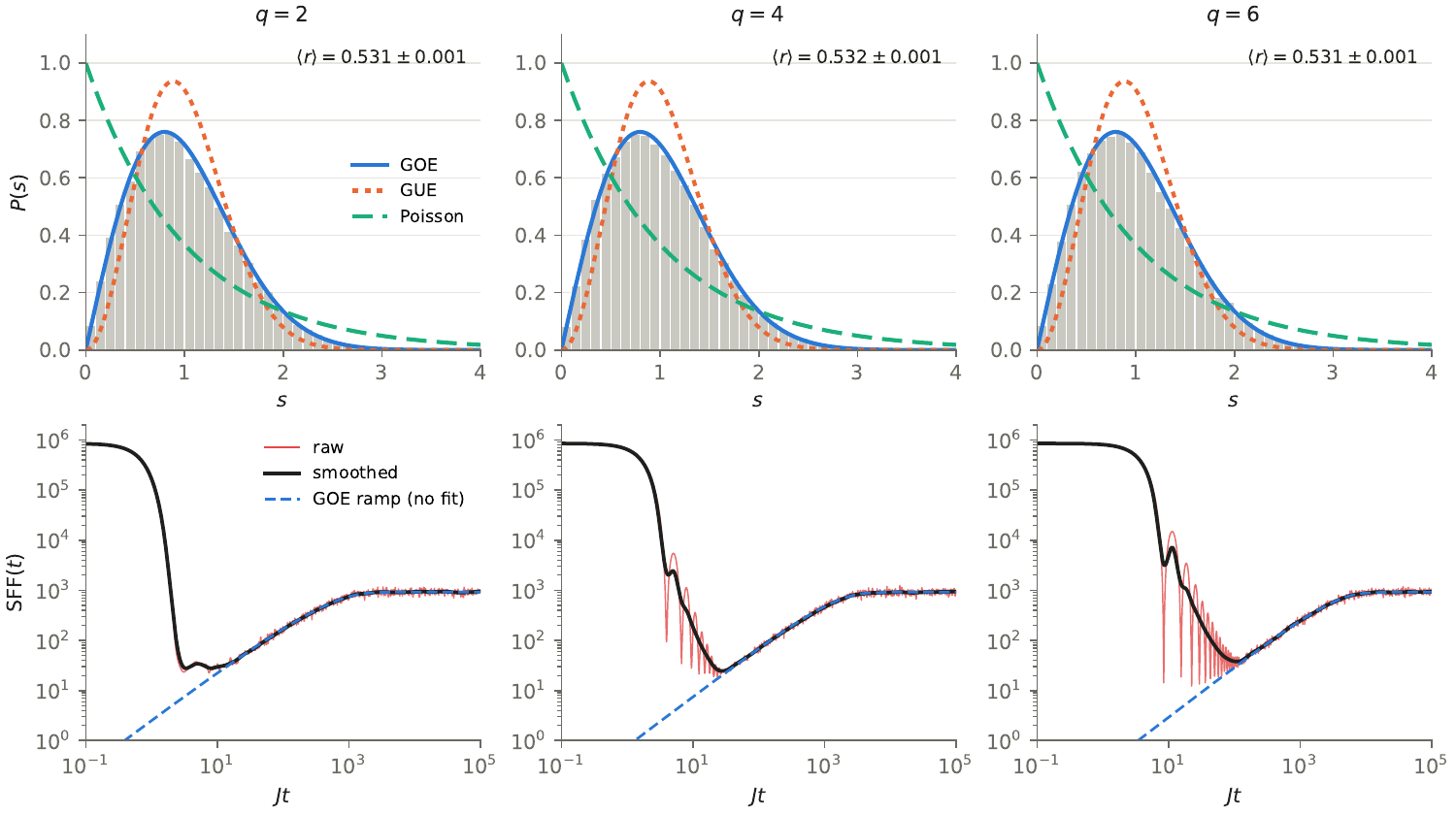}
    \caption{
    Quantum-chaos diagnostics at $N=12$, half filling, averaged over
    $R=100$ disorder realizations.
    \textbf{Top row:} nearest-neighbor level-spacing distribution $P(s)$
    from the middle $80\%$ of the spectrum, with spacings normalized to
    $\langle s\rangle=1$.
    \textbf{Bottom row:} disorder-averaged spectral form factor
    $\mathrm{SFF}(t)=\overline{|Z(it)|^2}$ for Hilbert-space dimension
    $\mathcal{D}=\binom{12}{6}=924$.
    The raw data (red) and Gaussian smoothing in logarithmic time (black) are
    shown; the dashed blue curve is the GOE ramp computed from the
    disorder-averaged density of states, with no free parameters.
    In the top row, the solid blue and dotted orange curves are the GOE and GUE Wigner surmises, respectively; the dashed green curve is the Poisson distribution $P(s)=e^{-s}$.
    }
    \label{fig:chaos}
\end{figure*}

\paragraph{Level-spacing statistics.}
For each realization we unfold the middle $80\%$ of the spectrum and normalize nearest-neighbor spacings to unit mean (details in App.~\ref{app:spectral}). We also quote the adjacent-gap ratio from the full spectrum, which avoids the need for spectral unfolding \cite{AddAtas2013}. Figure~\ref{fig:chaos} compares the resulting distribution with GOE, GUE, and Poisson reference curves. The mean adjacent-gap ratio is
\begin{table}[ht]
    \centering
    \caption{
    Mean adjacent-gap ratio $\langle r\rangle$ at $N=12$
    ($R=100$, full spectrum).
    Large-dimension reference values \cite{AddAtas2013}: GOE $0.5307$, GUE $0.5996$, Poisson $0.3863$.
    }
    \label{tab:gapratio}
    \begin{tabular}{@{}lccc@{}}
        \toprule
        & $q=2$ & $q=4$ & $q=6$ \\
        \midrule
        $n_\uparrow=N/2$ & $0.5311(9)$ & $0.5322(9)$ & $0.5305(11)$ \\
        $n_\uparrow=N/2+1$ & $0.5976(10)$ & $0.5989(10)$ & $0.5999(11)$ \\
        \bottomrule
    \end{tabular}
\end{table}
consistent with GOE statistics in the half-filled sector and GUE statistics one sector away, for all three values of $q$.

For generic complex couplings, unitary-class level statistics are expected in a chaotic sector unless an additional antiunitary symmetry acts within it. The non-overlap restriction gives the Hamiltonian the antiunitary symmetry $\mathcal A=PK$, where $P=\prod_{\mu=1}^{N}\sigma_\mu^x$ and $K$ is complex conjugation in the $\sigma^z$ basis. Its local action is
\begin{equation}
 \mathcal A(\sigma^x,\sigma^y,\sigma^z)\mathcal A^{-1}
 =(\sigma^x,\sigma^y,-\sigma^z),\qquad \mathcal A^2=1.
\end{equation}
It therefore preserves the half-filled sector and places it in the orthogonal symmetry class, for which GOE statistics are expected in the chaotic regime. This is analogous to the role of particle--hole symmetry in complex SYK \cite{KanazawaWettig2017,YouLudwigXu2017,GarciaGarciaVerbaarschot2016,Dowarah2026}. For the $q=4$ hard-core-boson model, GOE statistics at half filling and GUE statistics in other filling sectors were found in Ref.~\cite{AddFuSachdev2016}; the explicit form of $\mathcal A$ given here extends this to all even $q$. Away from half filling, $\mathcal A$ exchanges the sectors $M$ and $-M$ rather than acting within either one; a single such sector is generically in the unitary class.

Unlike conventional spin time reversal, $\mathcal A$ reverses only the longitudinal spin component. Conventional time reversal is $\Theta=(\prod_\mu i\sigma_\mu^y)K$, which reverses all three components and satisfies $\Theta^2=(-1)^N$. The two are related, up to an overall phase, by $\mathcal A\propto R_z(\pi)\Theta$, where $R_z(\pi)=e^{-i\pi M}$. Since the Hamiltonian has $U(1)$ symmetry, this rotation is itself a symmetry. For the even-$N$, half-filled clusters studied here, both antiunitary operators preserve the sector and square to $+1$.

\paragraph{Spectral form factor.}
The lower panels of Fig.~\ref{fig:chaos} show the characteristic finite-size slope--dip--ramp--plateau structure familiar from random-matrix descriptions of chaotic spectra and from SYK \cite{AddCotler2017,AddGarciaGarcia2021,AddOrman2025,Gharibyan2018}. We use the physical spectrum without unfolding or filtering and normalize the plotted SFF so that $\mathrm{SFF}(0)=\mathcal D^2$ and the plateau is of order $\mathcal D$; the black curve is obtained by Gaussian smoothing of $\log\mathrm{SFF}$ over approximately $0.15$ decade in logarithmic time. The ramp follows the parameter-free GOE prediction, which differs from the GUE one by about a factor of two at early times, so it provides additional support for the GOE assignment. The adjacent-gap ratio provides the cleaner finite-size diagnostic, while the SFF is a complementary consistency check.

\paragraph{Relation to spin ordering.}
The ordering diagnostic in Sec.~\ref{sec:glass} probes the ground state, whereas the level statistics above probe the bulk of the spectrum. The present data therefore show that transverse order in the ground state at $q=2$, and growing ground-state correlations at $q=4$, coexist in the same Hamiltonian with random-matrix bulk spectral statistics; they do not establish ordering and chaos at the same energy density. Spectral signatures of glassiness are expected instead at energies where metastable states proliferate, where the ramp of the spectral form factor is enhanced by their number \cite{WinerSFFSpinGlass2022}.

\subsection{Numerical magnetization transport}
\label{sec:transport}

\begin{figure*}[t]
    \centering
    \includegraphics[width=\textwidth]{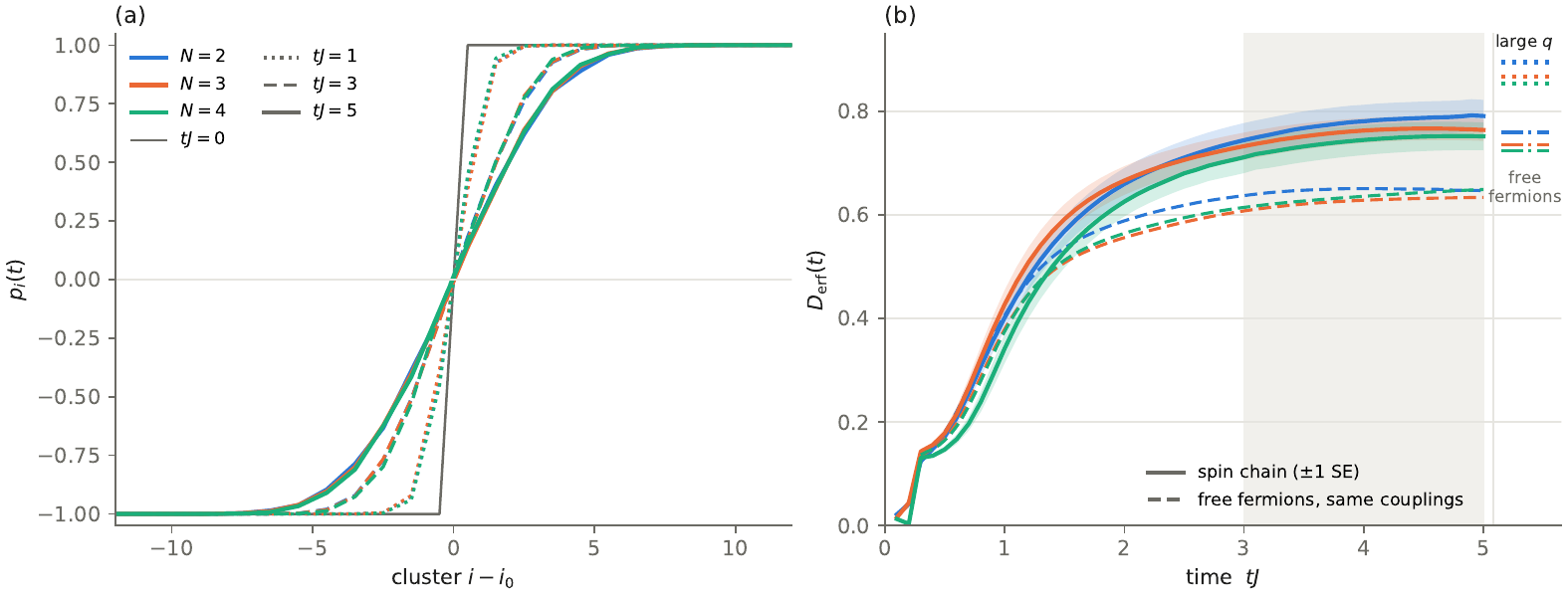}
    \caption{
    Domain-wall transport at $q=q'=2$ and $J=J'=1$.
    (a) Disorder-averaged polarization profiles $p_i(t)$; colors denote
    $N=2,3,4$ and line styles the displayed times.
    (b) $D_{\rm erf}(t)$ for the spin chain (solid; shading $\pm1$ disorder standard error) and for
    the free-fermion chain with the same couplings (dashed, $R=200$); the shaded band is the averaging
    window $3\leq tJ\leq5$. Short segments at the right give reference values evaluated at the finite-$N$
    bandwidth $\mathcal{J}_{\rm ch}(N)$ (Sec.~\ref{sec:transport}): the large-$q$ prediction
    $D=J'^2/\mathcal{J}_{\rm ch}=0.894$, $0.866$, $0.853$ at $\beta\to0$ (dotted) and the large-$N$
    free-fermion linear-response value $8J'^2/(3\pi\mathcal{J}_{\rm ch})=0.759$, $0.735$, $0.724$
    (dash-dotted) for $N=2,3,4$.
    The spin-chain data sets are $(L,\chi_{\rm MPS},R)=(48,128,100)$, $(48,128,48)$, and $(32,512,8)$ for $N=2,3,4$.
    }
    \label{fig:transport}
\end{figure*}

We probe the spreading of a fully polarized magnetization domain wall in the coupled-cluster chain with open boundaries, a standard probe of finite-temperature transport in one-dimensional lattice models \cite{Bertini2021May,LjubotinaZnidaricProsen2017}. We take $q=q'=2$ and $J=J'=1$. The $X^{(i)}$ are independent Hermitian complex-Gaussian matrices obeying Eqs.~\eqref{eq:hermiticity_X} and \eqref{eq:X_variance}; the $Y^{(i)}$ on different bonds are independent of each other and of the $X^{(i)}$, with independent proper complex-Gaussian entries of zero mean and variance $\overline{|Y_{\mu\nu}^{(i)}|^2}=1/(2N)$, as follows from Eq.~\eqref{eq:Y_variance}. Second-order TEBD with $U(1)$ conservation uses $\delta t=0.05/J$; for chains with $(N,L)=(2,6)$ and $(3,4)$ it agrees with exact evolution to $1.4\times10^{-4}$ in $p_i(t)$ up to $tJ=3$, with the error decreasing fourfold when $\delta t$ is halved. The data sets are $(N,L,\chi_{\rm MPS},R)=(2,48,128,100)$, $(3,48,128,48)$, and $(4,32,512,8)$, where $R$ counts disorder realizations.

We plot the cluster polarization, twice the magnetization density $m$ of Sec.~\ref{sec:limits},
\begin{equation}
    p_i(t)=\frac{2}{N}\langle M_i(t)\rangle
    =\frac{1}{N}\sum_{\mu=1}^{N}\langle\sigma^z_{\mu i}(t)\rangle .
\end{equation}
At each time, the profile is fit to $c+a\,\mathrm{erf}[(i-i_0)/w(t)]$, giving
\begin{equation}
    D_{\rm erf}(t)=\frac{w(t)^2}{4t}.
    \label{eq:D}
\end{equation}
Independently, the magnetization transferred across the center cut gives
\begin{equation}
    Q_{\rm tr}(t)=\left|\sum_{i<i_0}\left[p_i(t)-p_i(0)\right]\right|,
    \qquad
    D_{\rm tr}(t)=\frac{\pi Q_{\rm tr}(t)^2}{(\Delta p)^2t},
\end{equation}
With the initial jump $\Delta p=2$, averaging each realization over $3\leq tJ\leq5$ and then over disorder gives, for $N=2,3,4$,
\begin{align}
    D_{\rm erf}&=0.774\pm0.032,\quad 0.757\pm0.024,\quad 0.740\pm0.029,\nonumber\\
    D_{\rm tr}&=0.780\pm0.031,\quad 0.747\pm0.023,\quad 0.720\pm0.027.
\end{align}
Uncertainties are disorder standard errors of the per-realization window averages.

The profiles broaden symmetrically, the two estimators agree, and $D(t)$ depends only weakly on $N$, as expected from the normalizations in Eqs.~\eqref{eq:X_variance} and \eqref{eq:Y_variance}. The residual $N$ dependence follows the finite-$N$ bandwidth. The disjointness condition leaves $N(N-1)$ of the $N^2$ intracluster couplings at $q=2$, while the bond couplings are unconstrained, so $\mathcal{J}_{\rm ch}^2(N)=J^2(N-1)/(2N)+J'^2$; equivalently, the short-time check of App.~\ref{app:conductivity}, evaluated with the exact $q=2$ counts, gives $g''(0)=J^2(N-1)/N+2J'^2$ exactly. Normalized by the large-$q$ value at this bandwidth, $D_{\rm erf}\,\mathcal{J}_{\rm ch}(N)/J'^2=0.865$, $0.874$, and $0.868$ for $N=2,3,4$. Because $D(t)$ still drifts, the window averages are effective rather than asymptotic coefficients. Bond-dimension effects are small: on identical realizations, the window-averaged $D_{\rm erf}$ changes by $-0.0014\pm0.0005$ for $N=3$ between $\chi_{\rm MPS}=128$ and $256$, and by $-0.0067\pm0.0008$ and $-0.0018\pm0.0002$ for $N=4$ at $\chi_{\rm MPS}=256$ and $384$ relative to $\chi_{\rm MPS}=512$; $N=2$ is converged at $\chi_{\rm MPS}=128$. Stronger bond-dimension dependence at later times motivates the cutoff $tJ\leq5$.

At $q=q'=2$ the fermionic chain with the same couplings is special: it is free, and its single-particle states are Anderson localized in this quasi-one-dimensional geometry \cite{Wegner1979,FyodorovMirlin1991,Beenakker1997}, with band-center localization lengths of about $6.6$, $9.1$, and $11.7$ clusters for $N=2,3,4$. Its large-$N$ limit is nonetheless solvable. The local spectral function is a semicircle of variance $\mathcal{J}_{\rm ch}^2$, the total-current correlator reduces to the bond bubble for zero-mean random bond couplings (App.~\ref{app:conductivity}), and at $\beta=0$ each energy diffuses with $D(E)=\pi J'^2\rho(E)$, giving the linear-response value $D=8J'^2/(3\pi\mathcal{J}_{\rm ch})\approx0.693$ before localization sets in. This is smaller than the large-$q$ value $J'^2/\mathcal{J}_{\rm ch}\approx0.816$ of Sec.~\ref{subsec:limits-chain} by a factor $3\pi/8$. A domain wall superposes profiles with different $D(E)$; its transferred charge gives instead $D_{\rm tr}=\pi J'^2\big[\int\rho^{3/2}dE\big]^2\approx0.684$. Evaluated at the finite-$N$ bandwidth $\mathcal{J}_{\rm ch}(N)$ of the simulated chains, these formulas give $D=0.759$, $0.735$, $0.724$ and $D_{\rm tr}=0.749$, $0.726$, $0.714$ for $N=2,3,4$, and the large-$q$ value becomes $0.894$, $0.866$, $0.853$; these are the reference values shown in Fig.~\ref{fig:transport}(b). The spin chain is not free, and its large-$N$ diffusion constant at $q=2$ is not available in closed form.

Figure~\ref{fig:transport}(b) compares the two chains. In the window $3\leq tJ\leq5$ the spin chain is faster than the free-fermion chain at every $N$, by $0.11$ to $0.13$ (the free-fermion window averages are $0.648\pm0.021$, $0.626\pm0.014$, and $0.635\pm0.010$), a direct signature of the hard-core interactions. Both chains are still drifting upward at $tJ=5$, and their window averages lie $12$--$15\%$ below the corresponding finite-$N$ reference values of Fig.~\ref{fig:transport}(b): the large-$q$ value for the spin chain and the linear-response value for the free chain. We do not extrapolate to an asymptotic diffusion constant. Within the accessible window, extrapolated intercepts depend on the assumed form of the late-time approach at the $0.1$ level, and in one dimension long-time tails generically make $D(t)-D_\infty\propto t^{-1/2}$ rather than $1/t$ \cite{MukerjeeOganesyanHuse2006}; moreover, the free chain eventually localizes. The robust result is the finite-time enhancement over the free chain within the window. The comparison with the large-$q$ value is therefore qualitative: $q=2$ lies outside the large-$q$ regime, and the spin and fermion saddles already differ at the cubic local vertex (Sec.~\ref{subsec:limits-diff}).

The fully polarized wall samples every magnetization between $p=-1$ and $p=1$, so $D_{\rm erf}$ and $D_{\rm tr}$ could average a magnetization-dependent $D(m)$. We therefore repeated the $N=3$ calculation for weaker walls on the same disorder realizations ($L=24$, $R=48$), starting from Fock product states with a fixed number of up spins per cluster, which also have $\langle H\rangle=0$. A wall at half filling with $p_i(0)=\pm1/3$ gives the same $D_{\rm tr}$ as the full wall within $0.03$. This wall requires larger bond dimensions, because every bond is active from $t=0$: on eight realizations, $D_{\rm tr}$ shifts by $+0.035\pm0.006$ from $\chi_{\rm MPS}=128$ to $256$ and by $+0.014\pm0.002$ from $256$ to $512$, so the $\chi_{\rm MPS}=256$ value used for the comparison is estimated to be low by about $0.02$. This residual truncation-bias estimate is included in the quoted agreement, but does not establish full bond-dimension convergence. A wall confined to the nearly polarized region, $p_i(0)=-1/3\to-1$, is slower by $0.08\pm0.03$ in the same window. This is consistent with $D(m)$ decreasing toward full polarization, where the dynamics is that of dilute magnons and the one-magnon problem coincides with the free-fermion one. These weaker-wall results support the robustness of the finite-time spreading scale around half filling. The walls remain finite-amplitude perturbations of pure product states, however: $\langle H\rangle=0$ alone does not establish local equilibration, and these checks do not determine the linear-response diffusion coefficient \cite{Steinigeweg2017}.

These data do not settle the long-time behavior. In the free chain, localization reduces $D(t)$ appreciably only after $tJ\gtrsim20$ for $N=2$ and $tJ\gtrsim50$ for $N=4$, beyond the times reached here. Interacting SYK lattices can themselves undergo transitions between diffusive and localized or insulating behavior \cite{JianYao2017,AltlandBagretsKamenev2019}. Distinguishing asymptotic diffusion in the spin chain from slow localization therefore requires longer times than our tensor-network simulations can reach, or methods targeting late-time hydrodynamics \cite{Rakovszky2022}.

%%%%%%%%%%%%%%%%%%%%%%%%%%%%%%%%%%%%%%%%%%%%%%%%%
\section{Discussion and outlook}
\label{sec:outlook}

We discussed a $U(1)$-symmetric spin-SYK construction that retains the random all-to-all structure of complex SYK while working directly in a qubit Hilbert space. The model avoids the nonlocal Jordan--Wigner strings that arise when fermions are represented on qubits, and the physics of complex SYK remains accessible in the regime where the two share the large-$q$ saddle; at small $q$ the spin model orders, which marks where this correspondence ends.

The finite-size numerics identify two complementary features. First, the $U(1)$-invariant susceptibility gives strong evidence for transverse order in the ground state at $q=2$ and growing correlations at $q=4$, while no frozen order is visible at $q=6$ for the accessible sizes, all with $N<q^2$. Second, the bulk-spectrum level statistics follow random-matrix theory over the same range of $q$. These observations motivate a more detailed study of how low-energy ordering and potential glassiness depend on $q$, filling, and interaction sparsity, and how that crossover is related to the regime in which the large-$N$, large-$q$ saddle becomes quantitatively accurate.

Coupling the clusters introduces a conserved magnetization current. The controlled $q'=q$ large-$q$ calculation supplies an analytic conductivity and diffusion benchmark. The MPS calculation shows enhanced finite-time domain-wall spreading relative to the corresponding free-fermion chain. The accessible times do not determine an asymptotic diffusion constant. It remains to determine whether the spin chain exhibits asymptotic diffusion beyond the times reached here, where its free-fermion counterpart localizes, and whether operator growth fronts display the sharp-to-diffusive broadening predicted in related random-circuit and coupled-cluster settings \cite{AddNahum2018,AddVonKeyserlingk2018,KhemaniVishwanathHuse2018,RakovszkyPollmannVonKeyserlingk2018,XuSwingle2019,KhemaniHuseNahum2018,AleinerFaoroIoffe2016}. The particularly interesting low-temperature regime at finite $N$ and $q$ remains open.

%%%

The numerical methods used here already show clear limitations, and we expect most classical approaches to struggle at large $L$ and $N$. Exact diagonalization is restricted to a few tens of spins in total: at half filling, fixing the total magnetization reduces the Hilbert-space dimension $2^{NL}$ only by a factor of order $\sqrt{NL}$. Tensor-network methods face two distinct costs: grouping each cluster into one site gives an exponentially large local dimension $d=2^N$ \cite{Stolpp2021}, while entanglement between clusters sets the required MPS bond dimension. If intercluster entanglement grows linearly in time, maintaining accuracy generally requires an exponentially growing bond dimension over that regime. Entanglement within a grouped site does not itself increase the intersite bond dimension, and its dependence on $N$ must be assessed separately \cite{Schollwoeck2011Jan}. Entanglement bounds and quantum-advantage resource estimates for chains of spin-SYK dots coupled by bonds that do not transfer charge were recently obtained in Ref.~\cite{DeyCarrerasPatra2026}; the $U(1)$ chain studied here adds a conserved transported charge. Quantum Monte Carlo faces a sign problem from the random complex couplings, and real-time correlators would in any case require analytic continuation \cite{TroyerWiese2005}. Semiclassical methods such as the discrete truncated Wigner approximation \cite{Schachenmayer2015} have no small parameter for spin-$\tfrac12$, and their classical Lyapunov exponents are not constrained by the $2\pi T$ bound. Neural-network quantum states \cite{CarleoTroyer2017} must represent a random sign structure and optimize over a complex variational landscape. Finally, the large-$N$, large-$q$ analysis of Sec.~\ref{sec:limits} provides controlled benchmarks but does not address glassiness or the large-$L$ regime at finite $N$ and $q$. Classical methods are therefore best suited to small-instance benchmarks and high-temperature transport \cite{Zlokapa2026}, while low-temperature dynamics, scrambling, and glassy relaxation at large $L$ and $N$ remain difficult.

For quantum simulation, the most compelling next step is not simply to implement the dense model, but to identify observables for which analytically controlled limits provide benchmarks while finite-$N$, finite-$q$ dynamics remain classically challenging. Natural targets include conserved-charge transport, out-of-time-order correlators, low-energy thermodynamics, and the relation between ordering, glassiness, and scrambling. Ballistic operator spreading with a broadened front has been analyzed in random circuits and operator-hydrodynamic descriptions, including with a conserved charge, providing useful benchmarks for the Hamiltonian problem posed here \cite{AddNahum2018,AddVonKeyserlingk2018,KhemaniVishwanathHuse2018,RakovszkyPollmannVonKeyserlingk2018,AddHunterJones2018,XuSwingle2024,ChenGuLucas2020,AgarwalXu2022}. A sparsified spin-SYK model may further reduce the number of interaction terms \cite{Gautam2025Dec}, but the extent to which sparsification preserves the relevant large-$N$ saddle and finite-size chaotic dynamics is an important open question that is well suited to quantum simulation. Dense fermionic SYK can be sparsified very substantially while retaining important chaotic and holographic signatures above sparsity-dependent thresholds \cite{AddXuSparse2020,AddGarciaGarcia2021,AddOrman2025}, and sparse SYK has already motivated explicit quantum-simulation resource studies and trapped-ion experiments \cite{Tezuka2023Feb,Baumgartner2024,AddGranet2025}.

Finally, a realistic hardware study will require explicit circuit decompositions, resource estimates \cite{ChenHelsenOzols2026}, and an analysis of symmetry-breaking and decoherence. Thermofield-double preparation may provide one route to finite-temperature observables; at high temperatures, SYK Gibbs states can also be prepared efficiently by fast-mixing dynamics \cite{CaiZhanZlokapa2026}. Variational and adiabatic/parent-Hamiltonian approaches have been developed specifically for SYK and spin-SYK settings, while direct variational preparation of SYK thermal states has also been explored on quantum hardware \cite{AddSuTFD2020,AddArazThermal2024,AddKhorTFD2026}. The usefulness of these approaches here should be assessed together with the cost of state preparation and measurement.

\begin{acknowledgments}
We thank Jan C. Louw and Masanori Hanada for collaboration in the early stages of this work. J.C.H.~acknowledges funding by the Max Planck Society, the Deutsche Forschungsgemeinschaft (DFG, German Research Foundation) under Germany's Excellence Strategy---EXC-2111---390814868, and the European Research Council (ERC) under the European Union's Horizon Europe research and innovation program (Grant Agreement No.~101165667)---ERC Starting Grant QuSiGauge. This work is part of the Quantum Computing for High-Energy Physics (QC4HEP) working group. J.C.H.~acknowledges the hospitality of the Aspen Center for Physics, which is supported by National Science Foundation grant PHY-2210452, while this paper was being completed. We acknowledge support from the U.S. Department of Energy through GeoFlow DE-SC0019380 (S.G.) and from the Heising-Simons Foundation through grant 2024-4849 (B.S.). We also acknowledge the use of AI models, including GPT 6 Astra, Claude Opus 5.5, and Claude Fable 5.1, to carry out and check calculations and aid in writing the manuscript.
\end{acknowledgments}

\appendix

\section{\texorpdfstring{Details of the large-$N$ and large-$q$ saddle}{Details of the large-N and large-q saddle}}
\label{app:saddle-details}
This appendix collects the intermediate calculations supporting Sec.~\ref{sec:limits}.

\subsection{Disorder averaging and tuple counting}
\label{app:averaging}

Write $H=J\sum_{\bm\mu\bm\nu}X^{\bm\mu}_{\bm\nu}\mathcal{O}_{\bm\mu\bm\nu}$ with
$\mathcal{O}_{\bm\mu\bm\nu}=\prod_{a\in\bm\mu}\sigma^+_a\prod_{b\in\bm\nu}\sigma^-_b$,
so $\mathcal{O}^\dagger_{\bm\mu\bm\nu}=\mathcal{O}_{\bm\nu\bm\mu}$ because operators on distinct sites commute and Eq.~\eqref{eq:hermiticity_X} gives $H=H^\dagger$. Take the independent couplings $X$ to be proper complex Gaussian variables with $\overline{X^{\bm\mu}_{\bm\nu}X^{\bm\nu}_{\bm\mu}}=
\overline{|X|^2}$ and $\overline{(X^{\bm\mu}_{\bm\nu})^2}=0$. Averaging the time-ordered exponential,
\begin{equation}
    \overline{\mathcal{T}e^{-\int_0^\beta H}}
    =\mathcal{T}\exp\Big[\frac{J^2}{2}\,\overline{|X|^2}
    \sum_{\bm\mu\bm\nu}\int\!\!\int
    \mathcal{O}_{\bm\mu\bm\nu}(\tau_1)\mathcal{O}_{\bm\nu\bm\mu}(\tau_2)\Big].
\end{equation}
Because of the non-overlapping condition on the microscopic Hamiltonian, the product of operators factorizes over sites,
$\mathcal{O}_{\bm\mu\bm\nu}(\tau_1)\mathcal{O}_{\bm\nu\bm\mu}(\tau_2)
=\prod_{a\in\bm\mu}\sigma^+_a(\tau_1)\sigma^-_a(\tau_2)
 \prod_{b\in\bm\nu}\sigma^-_b(\tau_1)\sigma^+_b(\tau_2)$,
and, up to $O(q^2/N)$ corrections from restricting the sums to distinct, disjoint indices, the tuple sums become powers of the collective operators
$\hat G(\tau_1,\tau_2)=\frac1N\sum_a\sigma^+_a(\tau_1)\sigma^-_a(\tau_2)$
and $\hat G(\tau_2,\tau_1)$:
\begin{equation}
\begin{aligned}
    &\sum_{\bm\mu\bm\nu}\mathcal{O}_{\bm\mu\bm\nu}(\tau_1)\mathcal{O}_{\bm\nu\bm\mu}(\tau_2)
    \\
    &\qquad\simeq\Big[\frac{N^{q/2}}{(q/2)!}\Big]^{2}
    \hat G(\tau_1,\tau_2)^{q/2}\,\hat G(\tau_2,\tau_1)^{q/2}.
\end{aligned}
\end{equation}
Inserting the variance \eqref{eq:X_variance}, the prefactor collapses to
\begin{equation}
\begin{aligned}
    &\frac{J^2}{2}\,\overline{|X|^2}\Big[\frac{N^{q/2}}{(q/2)!}\Big]^{2}
    \hat G^{q/2}\hat G^{q/2}
    \\
    &\qquad=\frac{J^2N}{4q^2}\,
    \big[4\,\hat G(\tau_1,\tau_2)\hat G(\tau_2,\tau_1)\big]^{q/2}.
\end{aligned}
    \label{eq:prefactor}
\end{equation}
This is the role of the factor $2^{q-1}$ in Eq.~\eqref{eq:X_variance}: it converts the
$2^{-q}$ from $q$ projector traces into the natural variable $4G(\tau_1,\tau_2)G(\tau_2,\tau_1)$, and
the coefficient is fixed by
$\overline{\langle\!\langle H^2\rangle\!\rangle}=J^2N/2q^2$ to leading order in $q^2/N$ (the second-order
term of the exponential at equal times reproduces
$\tfrac12\langle\!\langle H^2\rangle\!\rangle$).

\subsection{Local chronological-ordering signs}
\label{app:local-signs}

For the four-point function only the eight chronological orderings in which
$\sigma^\pm$ alternate are nonvanishing. For each of them one checks that the
fermionic permutation sign equals $\sgn(\tau_{12})\sgn(\tau_{34})$; e.g.\
$\tau_1>\tau_4>\tau_3>\tau_2$ is a single transposition with
$\sgn(\tau_{12})\sgn(\tau_{34})=(+)(-)$. Hence
$\langle\mathcal{T}\sigma^+_1\sigma^-_2\sigma^+_3\sigma^-_4\rangle
=\sgn(\tau_{12})\sgn(\tau_{34})\,[\,G_f(1,2)G_f(3,4)-G_f(1,4)G_f(3,2)\,]$,
the disconnected parts cancel against $G(1,2)G(3,4)$ by Eq.~\eqref{eq:dict},
and the connected part is the fermionic bubble with the stated dressing.

However, the agreement stops at third order. For the six-point function the
chronological ordering $\tau_1>\tau_4>\tau_5>\tau_2>\tau_3>\tau_6$ is
alternating and nonvanishing; its permutation has four inversions (sign $+$),
whereas $\sgn(\tau_{12})\sgn(\tau_{34})\sgn(\tau_{56})=(+)(-)(+)=-$.
No pairwise dressing can repair this, because the two-point function already
fixes the dressing to $\pm\sgn$.

\subsection{Reduction to the large-q equation}
\label{app:largeq-steps}

Because $G^0(\tau)G^0(-\tau)=\nb(1-\nb)$ is $\tau$-independent,
\begin{equation}
\begin{aligned}
    &\big[4G(\tau_1,\tau_2)G(\tau_2,\tau_1)\big]^{q/2-1}
    \\
    &\qquad\to\big[1-4m^2\big]^{q/2-1}\,e^{\,g_s(\tau_1,\tau_2)},
    \\
    &g_s\equiv\tfrac12\big[g(\tau_1,\tau_2)+g(\tau_2,\tau_1)\big].
\end{aligned}
\end{equation}
Applying $(\partial_{\tau_1}+\hh)(-\partial_{\tau_2}+\hh)$ to
Eq.~\eqref{eq:linDyson} gives, for $\tau_1\neq\tau_2$,
$-G^0_f\,\partial_{\tau_1}\partial_{\tau_2}g/q=\Sigma_f$, and inserting
Eq.~\eqref{eq:EOM2},
\begin{equation}
    \partial_{\tau_1}\partial_{\tau_2}\,g(\tau_1,\tau_2)
    =-\,J^2\big[1-4m^2\big]^{q/2-1}e^{\,g_s(\tau_1,\tau_2)} .
\end{equation}
The right-hand side is symmetric, so the antisymmetric part of $g$ is a zero
mode $g_a\propto\tau_{12}$, which periodicity of $g$ (KMS for $G$, already
satisfied by $G^0$) sets to zero. With $g_s=g_s(\tau_{12})$ one arrives at Eq.~\eqref{eq:liouville}.

\subsection{Short-time normalization check}
\label{app:single-shorttime}

\paragraph{Independent check of the coefficient.}
At $\beta=0$ and $m=0$, $G''(0^+)=\langle\!\langle K K^\dagger\rangle\!\rangle$
with $K=[H,\sigma^+_1]$. Only monomials with $1\in\bm\nu$ contribute,
$[\mathcal{O}_{\bm\mu\bm\nu},\sigma^+_1]=-(\text{rest})\,\sigma^z_1$, and there are
$\binom{N-q/2}{q/2}\binom{N-1}{q/2-1} \sim \binom{N}{q/2}^{\!2}\,\tfrac{q}{2N}$ of them,
each with $\langle\!\langle(\text{rest})(\text{rest})^\dagger\rangle\!\rangle=2^{1-q}$.
Hence $\langle\!\langle KK^\dagger\rangle\!\rangle
\simeq J^2\,\overline{|X|^2}\binom{N}{q/2}^2\tfrac{q}{2N}\,2^{1-q}=J^2/2q$, and with
$G=\tfrac12(1+g/q)$ this gives $g''(0)=J^2=2\Jeff^2$, in agreement with
Eq.~\eqref{eq:liouville}. The factor $q/2N$ (rather than the Majorana $q/N$) is
the origin of the $\tfrac12$ in $\Jeff^2$.

\subsection{Thermodynamic integration}
\label{app:thermo-integration}

The fixed-$m$ free energy in Eq.~\eqref{eq:canonical-free-energy} follows from the saddle itself. Since $H\propto J$,
$\langle H\rangle/N=2J^2\partial_{J^2}f
=-\frac{J^2A(m)^{q/2}}{2q^2}\int_0^\beta d\tau\,e^{g_s(\tau)}$, and
Eq.~\eqref{eq:liouville} gives the integral as
$[g_s'(\beta)-g_s'(0)]/2\Jeff^2$, so that
$\langle H\rangle/N=-(2\pi v/\beta q^2)\,A(m)\tan(\pi v/2)$. This equals
$\partial_\beta(\beta f)$ for Eq.~\eqref{eq:canonical-free-energy}, because
$d\Phi_{\rm c}/dx=2\sin(\pi v/2)$, and integrating from $\beta=0$, where
$\Phi_{\rm c}=0$, fixes $f$. It agrees with the complex-SYK result
under Eq.~\eqref{eq:mapping-fixed} \cite{Louw2023Feb}; at $m=0$ it reduces to
twice the standard Majorana large-$q$ correction
\cite{AddMaldacenaStanford2016}.

\subsection{Coupled-cluster action and counting}
\label{app:chain-action}

Take $L$ clusters with fields $G_i$, and intercluster couplings
$Y^{(i\to i+1)}$ complex Gaussian with the variance
in Eq.~\eqref{eq:Y_variance}
(if $Y$ is taken real, the additional average $\overline{Y^2}$ generates
anomalous bilocals $\langle\sigma^+\sigma^+\rangle$, which vanish on the
$U(1)$-symmetric saddle and can be dropped). Averaging $H_{i\to i+1}+{\rm h.c.}$
using the disorder-averaging procedure of App.~\ref{app:averaging} gives the following result, where $G_i^{ab}\equiv G_i(\tau_a,\tau_b)$:
\begin{equation}
\begin{aligned}
    -S[\{G_i,\Sigma_i\}]
    =\sum_i\Big\{&\ln Z_{\rm loc}[\Sigma_i]+\int\!\!\int\Sigma_i G_i
    \\
    &+\frac{J^2}{4q^2}\int\!\!\int\big[4G_i^{12}G_i^{21}\big]^{q/2}
    \\
    &+\frac{J'^2}{2q'^2}\int\!\!\int
    \big[4G_{i+1}^{12}G_i^{21}\big]^{q'/2}\Big\}.
\end{aligned}
    \label{eq:chain-action}
\end{equation}
Note the coefficient $J'^2/2q'^2$ versus $J^2/4q^2$: the bond term contains
both $H_{i\to i+1}$ and its conjugate, and the two contributions coincide after
$\tau_1\leftrightarrow\tau_2$. This accounts for the relative factor of two between the bond and intracluster terms in Eq.~\eqref{eq:chain-action}.
Because the two tuple operators in an intercluster monomial act on different
clusters, no disjointness constraint arises for the bond term. Indices must
still be distinct within each tuple, so the power approximation carries a
factor $[(N)_{k}/N^{k}]^2=1-k(k-1)/N+O(N^{-2})$, with $k=q'/2$ and
$(N)_k=N(N-1)\cdots(N-k+1)$; this correction vanishes for $q'=2$.

The equations of motion are Eq.~\eqref{eq:EOM1} for each $i$ and
\begin{equation}
\begin{aligned}
    \Sigma_i(\tau_1,\tau_2)
    &=-\frac{J^2}{q}\big[4G_i^{12}G_i^{21}\big]^{q/2-1}G_i^{12}
    \\
    &\quad-\frac{J'^2}{q'}\sum_{s=\pm1}
    \big[4G_i^{21}G_{i+s}^{12}\big]^{q'/2-1}G_{i+s}^{12}.
\end{aligned}
\end{equation}

\section{\texorpdfstring{Large-$q$ conductivity}{Large-q conductivity}}
\label{app:conductivity}

We derive Eq.~\eqref{eq:largeq-conductivity} for $q'=q$. With
$\langle A;B\rangle^{R}(t)=i\theta(t)\langle[A(t),B]\rangle$, the spectral
representation gives
$\mathrm{Im}\,\langle I_{\rm tot};I_{\rm tot}\rangle^{R}(\omega)/L=\tfrac12(1-e^{-\beta\omega})S(\omega)\geq0$ for $\omega>0$,
where the total current $I_{\rm tot}=\sum_i I_{i,i+1}$ carries the response to
a uniform field gradient. Here $S(\omega)=\int dt\,e^{i\omega t}W(t)$ and
$W(t)=\langle I_{\rm tot}(t)I_{\rm tot}(0)\rangle/L$. Hence
\begin{equation}
    \sigma_{\rm dc}=\frac{\beta}{2}\int_{-\infty}^{\infty}dt\,W(t).
    \label{eq:app-kubo}
\end{equation}

\paragraph{Reduction to the bond bubble.}
Currents on different bonds are not dynamically uncorrelated: the time
evolution contains the same bond couplings, and a transfer across bond
$(i-1,i)$ followed by one across $(i,i+1)$ correlates the two currents at
order $t^2$ with a coefficient of order $N$. Single-bond and cross-bond
correlators are therefore separately of order $N$, and only their sum, $W$,
reduces to a simple form. To see this, couple a uniform imaginary-time gauge
field $a(\tau)$ to every bond, $H_{i\to i+1}\to e^{i(q/2)a(\tau)}H_{i\to i+1}$, so
that $W$ follows from the second derivative of $\ln Z$ with respect to $a$. In
the averaged action, Eq.~\eqref{eq:chain-action}, $a$ enters only through a
factor $e^{i(q/2)[a(\tau_1)-a(\tau_2)]}$ in each bond term. To first order in $a$,
the bonds $(i-1,i)$ and $(i,i+1)$ give equal and opposite sources for $G_i$ on the
homogeneous saddle, since cluster $i$ receives charge through one and emits it
through the other, so the saddle shifts only at $O(a^2)$. The $O(a^2)$ action is
then the second derivative of the bond terms at fixed $G$: its
$a(\tau_1)a(\tau_2)$ part is the bond bubble below, and its $a(\tau)^2$ part is the
contact term. Vertex corrections, which would enter through a linear shift of
the saddle, are absent for the total current at leading order in $1/N$,
although they are present for a single bond. For the free-fermion chain with
the same couplings we checked this directly at $\beta=0$: at $N=20$--$80$ the
single-bond and nearest-neighbor cross-bond correlators are separately of
order $N$, while their sum agrees with the bubble; the time integral of $W$
matches the bubble to $1\%$ up to $tJ\approx1.5$, after which a small negative
tail, whose amplitude decreases as $1/L$ in open chains of $L=16$--$48$
clusters, slowly reduces it.

\paragraph{Bond bubble.}
By Eq.~\eqref{eq:current}, the bubble is
$\tfrac{q^2}{4}\big(\langle H_{i\to i+1}H_{i\to i+1}^\dagger\rangle+\langle H_{i\to i+1}^\dagger H_{i\to i+1}\rangle\big)$
with the couplings of the two vertices paired and the rest of the dynamics
entering only through the saddle $G$. The terms
$\langle H_{i\to i+1}H_{i\to i+1}\rangle$ and $\langle H^\dagger_{i\to i+1}H^\dagger_{i\to i+1}\rangle$
are not forbidden by the global $U(1)$, since $H_{i\to i+1}$ is neutral, but they
vanish at this order because $\overline{Y^2}=0$ for proper complex Gaussian
couplings. The expectation value factorizes over sites as in
App.~\ref{app:averaging}. In imaginary time,
\begin{equation}
\begin{aligned}
    \big\langle H_{i\to i+1}(\tau)H_{i\to i+1}^\dagger(0)\big\rangle
    &=J'^2\,\overline{|Y|^2}\binom{N}{q/2}^{2}
    \\
    &\quad\times\big[G_{i+1}(\tau,0)\,G_i(0,\tau)\big]^{q/2},
\end{aligned}
\end{equation}
and likewise for the second ordering. On the homogeneous large-$q$ saddle of
Sec.~\ref{subsec:limits-chain}, $[G(\tau,0)G(0,\tau)]^{q/2}\to[A(m)/4]^{q/2}e^{g_s(\tau)}$,
and Eq.~\eqref{eq:Y_variance} gives
\begin{equation}
    W(\tau)=\frac{J'^2N}{4}\,A(m)^{q/2}\,e^{g_s(\tau)} .
    \label{eq:app-bubble}
\end{equation}
At $\tau=0$, $\beta=0$, and $m=0$ this is $J'^2N/4$. Since equal-time currents
on different bonds are uncorrelated at $\beta=0$, this equals
$\overline{\langle\!\langle I_{i,i+1}^2\rangle\!\rangle}$, which also follows
directly from Eqs.~\eqref{eq:current} and \eqref{eq:Y_variance}.

\paragraph{Real-time integral.}
$W(t)$ is analytic for $-\beta<\mathrm{Im}\,t<0$, so the integral in
Eq.~\eqref{eq:app-kubo} can be taken along $t-i\beta/2$, where
$W(t-i\beta/2)=\langle I_{\rm tot}(\tau)I_{\rm tot}(0)\rangle|_{\tau=\beta/2+it}/L$. From
Eq.~\eqref{eq:thermal-solution} with $\Jeff\to\mathcal{J}_{\rm ch}$,
\begin{equation}
    e^{g_s(\beta/2+it)}=\frac{\cos^2(\pi v/2)}{\cosh^2(\pi v t/\beta)},
\end{equation}
and $\int dt\,\mathrm{sech}^2(\pi vt/\beta)=2\beta/(\pi v)$. Therefore
\begin{equation}
    \frac{\sigma_{\rm dc}}{N}
    =\frac{J'^2}{4}A(m)^{q/2}\,\frac{\beta^2\cos^2(\pi v/2)}{\pi v}
    =\frac{\pi v}{4}A(m)\,\frac{(\Jpeff)^2}{\mathcal{J}_{\rm ch}^2},
\end{equation}
using $\beta\mathcal{J}_{\rm ch}=\pi v/\cos(\pi v/2)$ and
$(\Jpeff)^2=J'^2A(m)^{q/2-1}$. This is Eq.~\eqref{eq:largeq-conductivity}. For
$\beta\to0$, $\pi v\simeq\beta\mathcal{J}_{\rm ch}$, the real-time correlator is
$W(t)=\tfrac14J'^2NA(m)^{q/2}\,\mathrm{sech}^2(\mathcal{J}_{\rm ch}t)$, and
$\sigma_{\rm dc}/N=\beta J'^2A(m)^{q/2}/(4\mathcal{J}_{\rm ch})$; with
$\chi_m^{-1}\simeq4T/A(m)$ this gives $D=J'^2A(m)^{q/2-1}/\mathcal{J}_{\rm ch}$, as
stated in Sec.~\ref{subsec:limits-chain}.

\paragraph{Short-time check of $\mathcal{J}_{\rm ch}$.}
The normalization of $\mathcal{J}_{\rm ch}$ can be checked independently of the
saddle, as in App.~\ref{app:single-shorttime}. At $\beta=0$ and $m=0$,
$G''(0^+)=\langle\!\langle KK^\dagger\rangle\!\rangle$ with
$K=[H,\sigma^+_{\mu i}]$. Besides the intracluster monomials with $\mu\in\bm\nu$,
the monomials of $H_{i\to i+1}$ and of $H^\dagger_{i-1\to i}$ that lower spin
$\mu$ of cluster $i$ contribute; they are equal in number to the intracluster
ones at leading order and have the same variance with $J\to J'$. Each of the
two bonds therefore adds $J'^2$ to $g''(0)=J^2$, giving
$g''(0)=J^2+2J'^2=2\mathcal{J}_{\rm ch}^2$, in agreement with the homogeneous
saddle of Sec.~\ref{subsec:limits-chain}.

\section{Spectral analysis}
\label{app:spectral}

For the spacing distributions of Fig.~\ref{fig:chaos}, each realization's
levels in the middle $80\%$ of the spectrum are unfolded with a degree-$7$
polynomial fit to the cumulative level count, and the unfolded spacings are
normalized to unit mean. The adjacent-gap ratios of Table~\ref{tab:gapratio} use
the full spectrum without unfolding. The spectral form factor is computed from
the physical spectrum with $\mathrm{SFF}(0)=\mathcal D^2$; the smoothed curve
averages $\ln\mathrm{SFF}$ with a Gaussian in $\log_{10}t$ of full width at
half maximum $0.15$. The random-matrix ramp is
\begin{equation}
    \mathrm{SFF}_{\rm conn}(t)=\int dE\,\rho(E)\,K\!\left(\frac{t}{2\pi\rho(E)}\right),
\end{equation}
with $\rho(E)$ the disorder-averaged density of states normalized to
$\int\rho\,dE=\mathcal D$ and the GOE form factor
$K(\tau)=2\tau-\tau\ln(1+2\tau)$ for $\tau\leq1$ and
$K(\tau)=2-\tau\ln[(2\tau+1)/(2\tau-1)]$ for $\tau>1$. It has no free
parameters.

\bibliography{ref.bib}

\end{document}